\documentclass{emulateapj}    
\usepackage{epsfig}
\begin{document}

\title{Measuring Metallicities with HST/WFC3 photometry\footnotemark[1] }

\author{Teresa L. Ross\altaffilmark{2}, 
  Jon Holtzman\altaffilmark{2},
  Barbara J. Anthony-Twarog\altaffilmark{3},
  Howard Bond\altaffilmark{4}, 
  Bruce Twarog\altaffilmark{3}, 
  Abhijit Saha\altaffilmark{5},   
  Alistair Walker\altaffilmark{6} }

\footnotetext[1]{Based on observations made with the NASA/ESA Hubble Space
  Telescope, obtained at the Space Telescope Science Institute, which
  is operated by the Association of Universities for Research in
  Astronomy, Inc., under NASA contract NAS 5-26555. These observations
  are associated with program 11729 and 11664. }
\altaffiltext{2}{Department of Astronomy, New Mexico State
  University, P.O. Box 30001, MSC 4500, Las Cruces, NM 88003-8001, 
  emails: rosst@nmsu.edu, holtz@nmsu.edu}
\altaffiltext{3}{Department of Physics and Astronomy, University of
  Kansas, Lawrence, KS 66045-7582, USA, bjat@ku.edu, btwarog@ku.edu }
\altaffiltext{4}{Space Telescope Science Institute, 3700 San Martin
  Drive, Baltimore, Maryland 21218, bond@stsci.edu} 
\altaffiltext{5}{National Optical Astronomy Observatory, PO box 26732,
  Tucson, AZ 85726, USA  }
\altaffiltext{6}{Cerro Tololo Inter-American Observatory (CTIO), National
  Optical Astronomy Observatory, Casilla 603, La Serena, Chile email:
  awalker@ctio.noao.edu }

\begin{abstract}
We quantified and calibrated the metallicity and temperature
sensitivities of colors derived from nine Wide Field Camera 3
(WFC3) filters aboard the Hubble Space Telescope (HST) using Dartmouth
isochrones and Kurucz atmospheres models. The theoretical isochrone
colors were tested and calibrated against observations of five well
studied galactic clusters: M92, NGC 6752, NGC 104, NGC 5927, and NGC
6791, all of which have spectroscopically determined metallicities
spanning $-2.30 <$ \textrm{[Fe/H]} $< +0.4$.  We found empirical
corrections to the Dartmouth isochrone grid for each of the following
color magnitude diagrams (CMD) (F555W--F814W, F814W), (F336W--F555W,
F814W), (F390M--F555W, F814W)  and (F390W--F555W, F814W). Using the
empirical corrections we tested the accuracy and spread of the
photometric metallicities assigned from CMDs and color-color diagrams
(which are necessary to break the age-metallicity degeneracy). Testing
three color-color diagrams [(F336W--F555W),(F390M--F555W),(F390W--F555W),
 vs (F555W--F814W)], we found the colors (F390M--F555W) and
(F390W--F555W), to be the best suited to measure photometric
metallicities.  The color (F390W--F555W) requires much less
integration time, but generally produces wider metallicity
distributions, and, at very-low metallicity, the MDF from
(F390W--F555W) is $\sim$60$\%$ wider than that from (F390M--F555W).
Using the calibrated isochrones we recovered the overall cluster
metallicity  to within $\sim$ 0.1 dex in \textrm{[Fe/H]} when using
CMDs (i.e. when the distance, reddening and ages are approximately
known).  The measured metallicity distribution function (MDF) from
color-color diagrams show this method measures metallicities of
stellar clusters of unknown age and metallicity with an accuracy of
$\sim$ 0.2 - 0.5 dex using F336W--F555W, $\sim$0.15 - 0.25 dex using
F390M--F555W, and $\sim$0.2 - 0.4 dex with F390W--F555W, with the
larger uncertainty pertaining to the lowest metallicity range.

\end{abstract}


\section{Introduction}

Metallicity, age and mass are fundamental characteristics of a stellar
population. Metallicity distributions, in conjunction with chemical
evolution models, provide evolutionary information about both
enrichment and gas inflow and outflow in the form of galactic
winds. While in recent years measuring large numbers of spectroscopic
metallicities has become more feasible with multiobject spectrographs
\citep{2011ApJ...727...78K},  it is still challenging to observe
samples large enough to find rare objects or substructure in
metallicity distribution functions, or to observe faint enough to
build up large samples in nearby galaxies. Photometric metallicities,
though not as accurate as spectra, provide measurements for every star
in the field, including those with fainter magnitudes than can be
reached spectroscopically.

The general technique to assign photometric metallicities relates 
color to metallicity. For instance, in a color magnitude diagram (CMD)
fiducial ridgelines from several clusters with similar age, and a
range of known metallicities can be interpolated to estimate the
metallicity of an unmeasured cluster based upon the location of its
ridgeline, provided that the cluster is also of similar age (e.g.,
Saviane et al. 2000; Da Costa et al. 2000,2002). 
\nocite{2000A&A...355..966S,2000AJ....119..705D,2002AJ....124..332D} 
Fiducial ridgelines have been used to derive empirical relations
between color and metallicity at a given absolute magnitude 
\citep{1990AJ....100..162D,2007ApJ...670..400C}. Alternatively, one
can use theoretical isochrones, however, the isochrones need to be
empirically calibrated to match observed sequences (e.g., Brown et
al. 2005; Lianou et al. 2011).
\nocite{2005AJ....130.1693B,2011A&A...531A.152L}   

An issue with deriving metallicities from color arises because a giant
can be redder either because it is older, or because it is more metal
rich.  One method to break the age-metallicity degeneracy uses a
color-color diagram, where one color is constructed using a
metallicity sensitive filter, and the other color is constructed from
a pair of filters that provide a temperature estimate with minimal
dependence on metallicity.

There is a long history in astronomy of using specifically designed
filters to isolate stellar characteristics such as metallicity. 
\citet{1966ARA&A...4..433S} proposed photometric indices as a means
of stellar classification. The $m_1 = (v-b)-(b-y)$ index estimates the
stellar metallicity of horizontal branch, red giant branch and main
sequence stars when used in conjunction with a temperature-sensitive
index such as $(b-y)$. The Washington system was developed to measure
photometric metallicities, temperatures and the amount of CN line
blanketing for giant G and K stars \citep{1976AJ.....81..228C}.

\begin{figure*}[ht]
\centering
  \epsfig{file=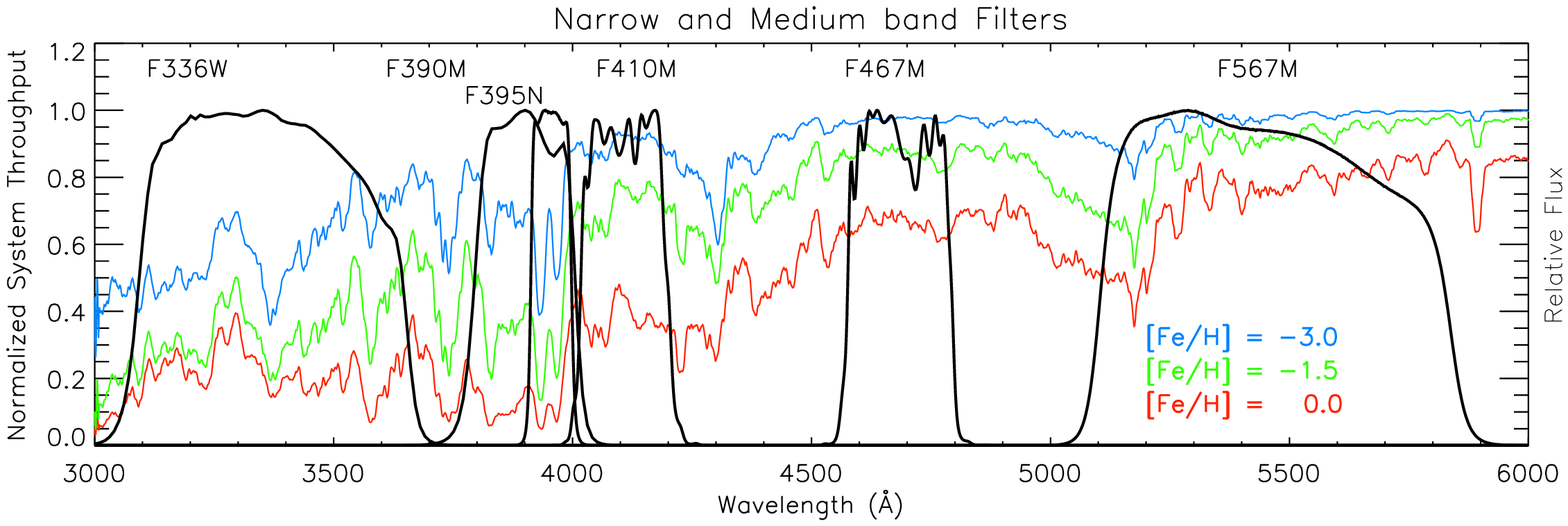,height=2.in }
  \epsfig{file=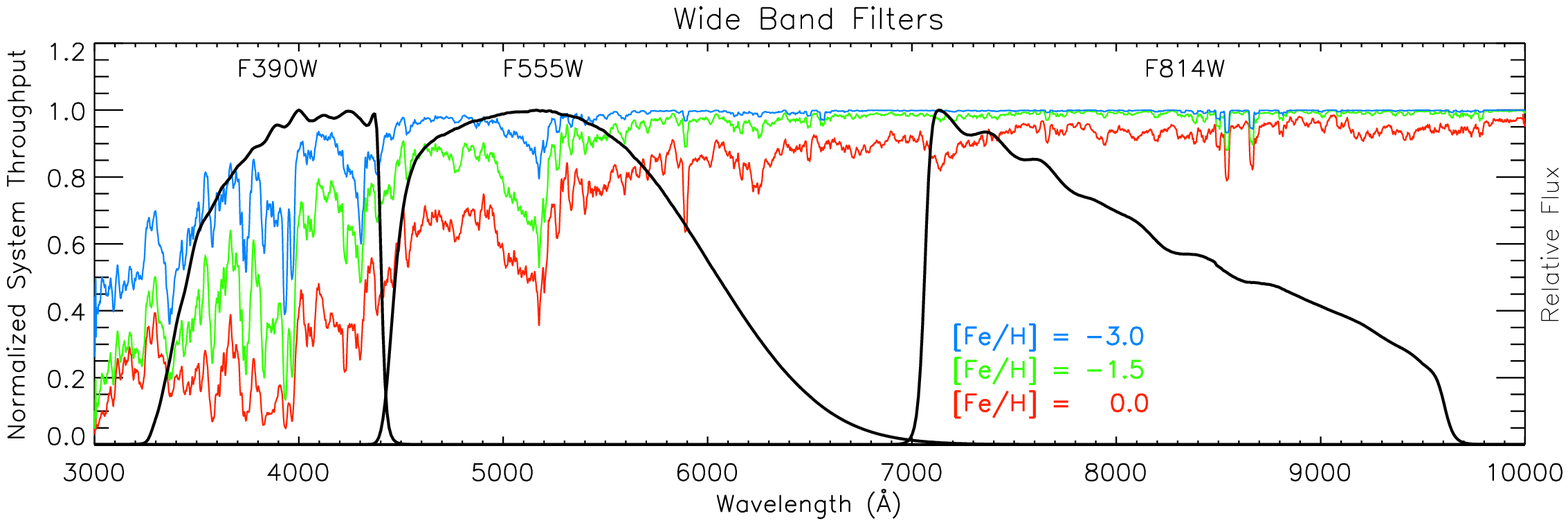,height=2.in }
\caption{UVIS filter transmission curves overlaid with Kurucz model
  stellar spectra of typical giant branch stars from log g = 1.5, T =
  4000 and \textrm{[Fe/H]} = 0.0, -1.5, $\&$ -3.0. (A color version of
  this figure is available in the online journal.) \label{bandpasses}}
\end{figure*}

Another consideration with photometric metallicities is that medium
and broad band colors lose nearly all sensitivity at low
metallicity. The Caby system, developed in the 1990's, modified the 
Str\"{o}mgren system, replacing the v filter with a filter centered on
the Ca H $\&$ K lines. The Caby system measures stellar metallicities
3 times more accurately, especially at low metallicity where the m$_1$
index loses sensitivity \citep{1991AJ....101.1902A}.  

Several filters on the HST/WFC3 were designed to provide information
on the metallicities of resolved populations. Our WFC3 calibration
program (11729, PI Holtzman) \nocite{2008hst..prop11729H} was designed
to collect images of clusters with well established spectroscopic
metallicities in order to map WFC3 colors to metallicity.  Another
program designed to study the galactic bulge (11664, PI Brown)
imaged the same clusters with different filters, also as a
calibration.  This paper utilizes observations from both programs to 
present a broad range of calibrating filters and their metallicity
sensitivities. Observations were obtained of M92, NGC 6752, NGC 104,
NGC 5927, and NGC 6791. The five calibration clusters are well studied 
spectroscopically, and span a wide range of metallicity: $-2.3 <$ 
\textrm{[Fe/H]} $< +0.4$.

Individual stars can be observed throughout the Local Group, opening 
up possibilities for studying populations within the Milky Way, as
well as Local Group galaxies. Our primary interest is in using these
calibrations to measure metallicity distribution functions in several
Local Group dwarf galaxies \citep{2009hst..prop12304H} where we only
have sufficient accuracy to measure giants, but the calibrations
presented here may have broader applicability. As will be shown,
photometric indices that measure metallicity also have a sensitivity
to surface gravity, but for objects outside the Milky Way, it is
trivial to separate giants from dwarfs based on their observed
brightness.

We structure the paper as follows: in Section 2 we use model
atmospheres and isochrones to explore the capacity of various filter
combinations to measure metallicity; in Section 3 we present the
observations of the stellar clusters. Section 4 describes how we
calibrate a set of isochrones to the observed sequences, and
demonstrate how well these can be used to recover metallicities. In
Section 5 we summarize our conclusions.

\section{Deriving Metallicity}

\subsection{WFC3 filters}
HST WFC3 observations were obtained in the following filters: F336W,
F390M, F390W, F395N, F410M, F467M, F547M, F555W, F814W, F110W
and F160W. Information on the filter widths and system throughputs are
listed in Table \ref{throughput}.  The system responses for the UVIS
filters are shown in Figure \ref{bandpasses}, along with Kurucz model
stellar spectra of typical giant branch stars with log g = 1.5, T =
4000 K and three different metallicities, \textrm{[Fe/H]} $= 0.0$,
$-1.5$ and $-3.0$.

\begin{deluxetable}{llrcl}[b]
\tablefontsize{\tiny}
\tablecaption{Filter data}
\tablehead{
     \colhead{Filter}  &
     \colhead{Description }&
     \colhead{Width }&
     \colhead{Peak system  }&
     \colhead{R$_{filter}$ }\\
	 &&(nm)& Throughput&
     }
\startdata
F336W  &u, Str\"{o}mgren   &51.1   &0.20  &5.04\\
F390W  &C, Washington      &89.6   &0.25  &4.47\\
F390M  &Ca II continuum    &20.4   &0.22  &4.63\\
F395N  &Ca II, 3933/3968   & 8.5   &0.22  &4.58\\
F410M  &v, Str\"{o}mgren   &17.2   &0.27  &4.42\\
F467M  &b, Str\"{o}mgren   &20.1   &0.28  &3.79\\
F547M  &y, Str\"{o}mgren   &65.0   &0.26  &3.12\\
F555W  &WFPC2 V            &156.2  &0.28  &3.16\\
F814W  &WFPC2 Wide I       &153.6  &0.23  &1.83\\
F110W  &Wide YJ            &443.0  &0.56  &1.02\\
F160W  &WFC3 H             &268.3  &0.56  &0.63
\enddata
\tablecomments{Widths and throughputs were taken from the WFC3
  Instrument Handbook. Widths listed are passband rectangular width,
  defined as the equivalent width divided by the maximum throughput
  within the filter bandpass,
  $\int$[T($\lambda$)d$\lambda$/max(T($\lambda$))]. 
  Calulations of R$_{filter}$ are described in Section 3.4.
\label{throughput} }
\end{deluxetable}

Many of these filters are comparable to those from well established
systems. For example, F336W, F410M, F467M and F547M are analogous to
Str\"{o}mgren  u, v, b and y, respectively.  While designed for
measuring the Balmer decrement, the F336W and F410M filters both cover
spectral regions with many absorption features from metals. The other
two Str\"{o}mgren filters, F467M and F547M, sample regions mostly
clear of spectral features; historically F467M--F547M (i.e. b--y) has
been used as  a temperature indicator.  

Filters F390M and F395N cover the Ca H $\&$ K spectral features. The
F395N filter is narrow (85 \AA ) while the F390M filter is broader,
leading to a higher throughput, but also including a CN feature at
$\lambda$ $\approx$ 3885 \AA  .  

The F390W filter has a wide bandpass (896 \AA ), and is similar to the
ground-based Washington C filter.  The Washington C filter was
designed to evaluate the total effects of line blanketing by CN (bands
at 3595, 3883 and 4215 \AA) as well as the CH molecular transition at 
4304 \AA, commonly known as the G band \citep{1976AJ.....81..228C}. 

F555W and F814W are wide band filters designed to cover the same
spectral regions as the WFPC2 and ACS filters of the same name; they
are similar to, but broader than, Johnson-Cousins V and I. The F555W
and F814W filters measure mostly continuum; one notable exception is
the MgH feature at $\sim$ 5100 \AA$\ $in the F555W bandpass.

As part of the observation campaign images in two near IR filters,
F110W and F160W, were obtained in an effort to explore reddening free
indices \citep{2009AJ....137.3172B}. However, we found that the
photometric uncertainties from a two color reddening free index seemed
to be too large to significantly add to our present analysis.

\subsection{Age - Metallicity Degeneracy}
Stellar colors are a function of gravity, metallicity 
and effective temperature. For a given star, increasing the
metallicity lowers the  effective temperature and enhances line
blanketing effects at a given mass, both of which cause redder
color. For populations of comparable age the color is directly related
to the metallicity. This relation has been used extensively in older
stellar populations, e.g., globular clusters
\citep{1990AJ....100..162D,2000A&A...355..966S}, to determine 
metallicities.

For an old population, we calculate the metallicity sensitivity of CMD
colors using Dartmouth stellar isochrones \citep{2008ApJS..178...89D}. 
These cover the metallicity range $-2.5 <$ \textrm{[Fe/H]} $< 0.5$. Since we
wish to understand sensitivity to metallicity at lower metallicities,
we extend these down to \textrm{[Fe/H]} $= -5$ by assuming that stars with
\textrm{[Fe/H]} $< -2.5$ have the same effective temperatures and luminosities
as those with \textrm{[Fe/H]} $= -2.5$, while adopting the colors from a grid
of model atmospheres \citep{2004astro.ph..5087C} that extends down to
\textrm{[Fe/H]} $= -5$. 

In Table \ref{cmdsens}, we report metallicity sensitivity for RGB
stars, in units of dex of \textrm{[Fe/H]} per 0.01 mag of color change, for a
range of different color choices, including most of the wideband UVIS
WFC3 filters. In these units, small numbers represent higher
sensitivity to metallicity. These sensitivities were computed for
stars at M$_{F814W}=-1$; for more luminous giants, the sensitivity is
better, while for fainter ones, it is worse. Sensitivities are
reported for several different ranges in metallicity, demonstrating
the smaller color sensitivity at lower metallicity. 

Generally, sensitivity increases as the wavelength separation of the 
filters increases. However, each filter has different photometric
precision for a fixed exposure time.  The second to last column in
Table \ref{cmdsens} gives the relative color errors (normalized to
$\sigma_{F555W-F814W}$), estimated using the exposure time calculator
(ETC) on the WFC3 online data handbook for a K5 III giant and a fixed 
exposure time.  The last column of Table \ref{cmdsens} lists the
sensitivity, over the metallicity range $-1.5 <$ \textrm{[Fe/H]} $<-0.5$,
scaled by the relative photometric error. 

The implication is that ideal filter choices rely on both the color
sensitivity to metallicity and the precision to which the color can be
measured. The most metallicity sensitive colors will not be the
optimum choice when observing time and errors are considered. However,
if you use a color that only changes minimally with metallicity, no
amount of increased photometric accuracy will improve the metallicity
determination.  

The last column of Table \ref{cmdsens} suggests that F475W--F814W is
the optimal color for maximum metallicity sensitivity, and this has
been adopted by many studies (e.g., Gallart 2008);
\nocite{2008ASPC..390..278G} although this choice does depend to some
extent on the color of the target stars. Other commonly used
metallicity sensitive colors are F390W--F814W and F555W--F814W. 
As pointed out in the last paragraph there is a trade off between
metallicity sensitivity and photometric accuracy (from a reasonable
amount of integration time).  The sensitivities in Table \ref{cmdsens}
show that these wide band colors have simultaneously greater
throughput and less metallicity sensitivity than the medium and narrow
band filters listed in Table \ref{cmdsens}. Additionaly, at very-low
metallicity, these broad-band colors have little sensitivity ($> 0.65$
and $> 1$ dex, respectively, per 0.01 mag of color change), and since
it is challenging to reduce photometric errors below 0.01 mag, this
leads to a fundamental limit on the accuracy of derived metallicities.
For some colors, at low metallicity the color change from metallicity
becomes smaller than the typical photometric accuracy of 0.01
mag. In Table \ref{cmdsens} we report this as $>$ 1 dex of \textrm{[Fe/H]}
/0.01 mag color change. The narrower filters, such as F395N and F390M,
retain sufficient sensitivity ($<0.15$ dex per 0.01 mag of color
change) to provide useful metallicity estimates even at very-low
metallicity.

\begin{deluxetable*}{lccccccc}
\tablecaption{CMD sensitivity: metallicity per 0.01 magnitude color change}
\tablehead{
\multicolumn{1}{l}{}&
\multicolumn{5}{c}{Sensitivities}&
\multicolumn{1}{l}{}&
\multicolumn{1}{l}{}\\
     \colhead{Color}&
     \colhead{-4.5 $<$}&
     \colhead{-3.5 $<$}&
     \colhead{-2.5 $<$}&
     \colhead{-1.5 $<$}&
     \colhead{-0.5 $<$}&
     \colhead{normalized}&
     \colhead{\textrm{[Fe/H]} / } \\
       \colhead{}  &
       \colhead{\textrm{[Fe/H]}}&
       \colhead{\textrm{[Fe/H]}}&
       \colhead{\textrm{[Fe/H]}}&
       \colhead{\textrm{[Fe/H]}}&
       \colhead{\textrm{[Fe/H]}}&
       \colhead{$\sigma_{color}$ }  &
       \colhead{(0.01 mag / }  \\
         \colhead{}  &
         \colhead{$< -3.5$}&
         \colhead{$< -2.5$}&
         \colhead{$< -1.5$}&
         \colhead{$< -0.5$}&
         \colhead{$<  0.5$}&
         \colhead{} &
         \colhead{ $\sigma_{color}$ ) } \\
\multicolumn{1}{l}{}&
\multicolumn{5}{c}{(dex \textrm{[Fe/H]}/0.01 mag of color change)}&
\multicolumn{1}{l}{}&
\multicolumn{1}{l}{}    }

\startdata
 F336W--F814W   &0.270  & 0.100   &0.021   & 0.007   & 0.005   & 44.2   & 0.31    \\
 F390W--F814W   &0.667  & 0.185   &0.036   & 0.012   & 0.009   &  4.5   & 0.05    \\
 F390M--F814W   &0.454  & 0.133   &0.028   & 0.008   & 0.007   & 31.3   & 0.26    \\
 F395N--F814W   &0.333  & 0.094   &0.027   & 0.010   & 0.007   & 69.7   & 0.72    \\
 F410M--F814W   & $>$1  & 0.476   &0.060   & 0.014   & 0.009   & 11.6   & 0.17    \\
 F438W--F814W   & $>$1  & 0.400   &0.053   & 0.020   & 0.013   &  2.9   & 0.06    \\
 F467W--F814W   & $>$1  & $>$1    &0.149   & 0.038   & 0.017   &  3.6   & 0.13    \\
 F475W--F814W   & $>$1  & $>$1    &0.097   & 0.031   & 0.017   &  1.3   & 0.04    \\
 F547M--F814W   & $>$1  & $>$1    &0.198   & 0.050   & 0.023   &  1.4   & 0.07    \\
 F555W--F814W   & $>$1  & $>$1    &0.182   & 0.048   & 0.022   &  1.0   & 0.05    \\
 F606W--F814W   & $>$1  & $>$1    &0.284   & 0.075   & 0.027   &  0.79  & 0.06    \\
 F625W--F814W   & $>$1  & $>$1    &0.402   & 0.114   & 0.033   &  0.90  & 0.10    \\
 F775W--F814W   & $>$1  & $>$1    & $>$1   & 0.741   & 0.270   &  0.94  & 0.70   
\enddata	  
\tablecomments{This table presents the various CMD color sensitivities
  to metallicity. The color difference is measured at $M_{F814W}$ = -1
  for metallicity spacing 1 dex of \textrm{[Fe/H]}. Sensitivity is defined here
  as dex of \textrm{[Fe/H]} / 0.01 mag of color spanned. The $\sigma_{color}$
  photometric errors were estimated using the WFC3 ETC and normalized
  and to $\sigma_{(F555W-F814W)}$.  The last column reports the
  \textrm{[Fe/H]} dex /(0.01 mag/$\sigma$) for the metallicity range $-1.5 <$
  \textrm{[Fe/H]} $<-0.5$.  At extremely low metallicities some of the color
  changes are beyond typical  photometric accuracy, i.e. greater than
  a dex of color / 0.01 mag of color change.    \label{cmdsens} } 
\end{deluxetable*}

 \begin{figure}[ht]
  \epsfig{file=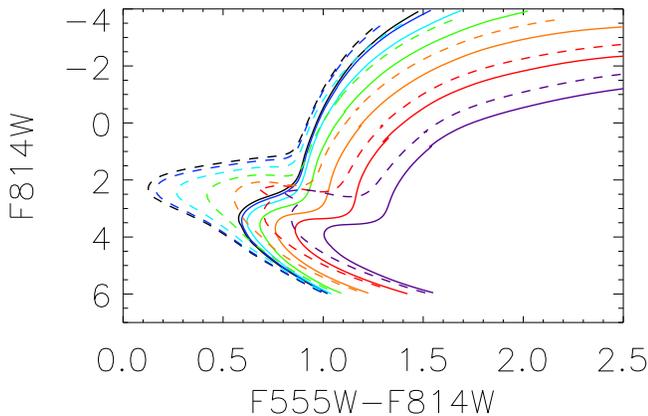, height=2.4in}
\caption{The age-metallicity degeneracy is shown here with isochrones
  of different ages covering a range of  metallicity: $ -2.5 <$
  \textrm{[Fe/H]} $< +0.5$, black represents the most 
  metal poor, purple the most metal rich, with each color in between
  representing a 0.5 increment in \textrm{[Fe/H]}; solid lines represent an age
  of 12.5 Gyr and the dashed lines represent 4 Gyr. 	\\	(A color
  version of this figure is available in the online journal.) \label{agez}} 
\end{figure}

\begin{figure}[ht]
  \epsfig{file=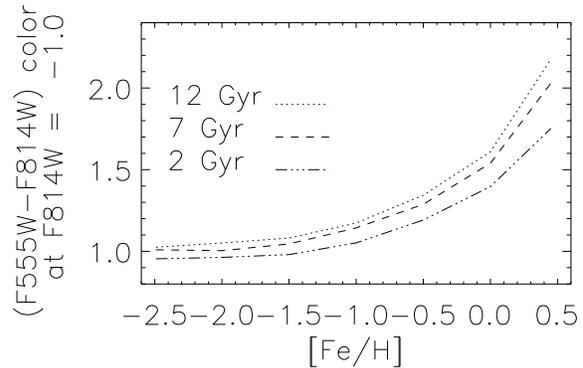, height=2in}
	\caption{Shown above are the giant branch (F555W--F814W) colors at
      fixed magnitude of F814W $=-1$ for populations of different ages
      (2, 7, and 12 Gyr) as a function of metallicity. This plot
      quantitatively demonstrates that small relative changes of color
      at low metallicity can lead to large changes in metallicity, and
      the effect is larger in older populations. \label{age_col}}   
\end{figure}

For a population of mixed age, the color-metallicity relation breaks
down because younger giants, which are more massive, are hotter than
older giants of the same metallicity.   The color changes due to age
and metallicity are demonstrated in Figure \ref{agez}, for two
different ages (12.5 and 4 Gyr, solid and dashed lines, respectively),
and for a range of metallicities (different colors). The effect is
quantitatively shown in Figure \ref{age_col}, where color as a
function of metallicity is plotted for several ages (2, 7 and 12 Gyr)
for a giant branch star with a $M_{F814W}$ = $-1.0$ (with comparable
results along the entire giant branch). At higher metallicity this
leads to an uncertainty in derived metallicity of a few tenths of a
dex, but the uncertainty can be significantly larger at lower
metallicity. The uncertainties in metallicity also become greater
with younger populations; as Figure \ref{age_col} shows, the color
difference between 2 and 7 Gyr is almost double that between 7 and 12
Gyr.

\subsection{Separating atmospheric parameters using photometric colors}

The key to breaking the age-metallicity degeneracy is separating the
color effects of metallicity and temperature.  Two color indices have
the potential to break the degeneracy because one color can measure a
metallicity dependent feature in the spectrum, while the other can
control for the temperature.  The more the two color's sensitivities
to metallicity and temperature differ, the more effective the color
combination will be at measureing metallicity.

\begin{figure}[h]
  \epsfig{file=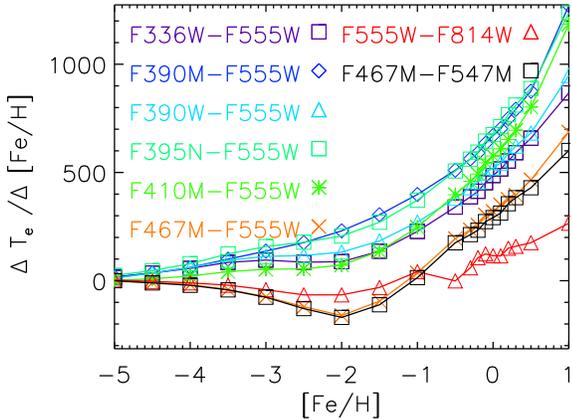,height=2.5in }
\caption{Relative color change from temperature compared to color
  change from metallicity, as a function of metallicity, i.e. ($\Delta$
  color / $\Delta$ \textrm{[Fe/H]}) / ($\Delta$ color/$\Delta$ T$_e$) ),
  measured from Kurucz atmospheric models. (A color version of
  this figure is available in the online journal.)  \label{dtdz}}
\end{figure}

We examine the temperature and metallicity sensitivity for all the
filter combinations by using  Kurucz stellar atmosphere models of
metallicity ranging from $-5 >$ \textrm{[Fe/H]} $> +1$. We integrate
the WFC3 transmission curve for each filter over the synthetic spectra
with parameters of typical giant branch stars (log g = 2.5 and
T$_{eff}$ = 4500 K).  We compute the color change with temperature
($\Delta$ color/$\Delta$ T$_e$) and metallicity ($\Delta$ color /
$\Delta$ \textrm{[Fe/H]}), at a range of metallicities. The ratio of
the two gives the relative sensitivity to temperature and metallicity
at equal color difference, where small values of $\Delta$
T$_e$/$\Delta$ \textrm{[Fe/H]} indicate a smaller dependence on
metallicity than temperature. The relative sensitivity of temperature
to metallicity ($\Delta$ T$_e$/$\Delta$ \textrm{[Fe/H]}) for all
colors is plotted in Figure \ref{dtdz}. As expected, the colors whose
bandpasses contain fewer metal-features (e.g F555W--F814W and
F467M--F547M) are the ones that are least sensitive to metallicity.

Figure \ref{dtdz} shows that the F467M--F547M and F555W--F814W colors
have comparably smaller sensitivities to metallicity.  However, the
relative color error for F467M--F547M is over 3.5 times larger than
F555W--F814W for a fixed exposure time. F555W--F814W stands out as the
optimal temperature index, considering minimal metallicity sensitivity
and photometric accuracy.  While F555W--F814W certainly changes with
metallicity (cf. its use as a metallicity indicator in clusters at
fixed age as discussed in the previous section) it has the largest
relative sensitivity of temperature to metallicity of all the filters
compared in this section (see Figure \ref{dtdz}).  

\begin{figure}[h]
  \epsfig{file=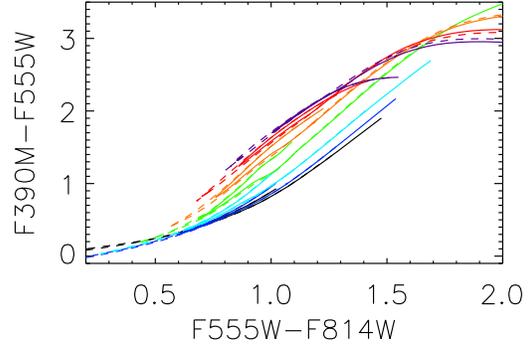,height=2.in }
	\caption{The color-color diagram breaks the age-metallicity
      degeneracy. Isochrones cover a range of metallicity: $-2.5 <$
      \textrm{[Fe/H]} $< +0.5$, the lowest (black) line represents the
      most metal poor, highest (purple) line the most metal rich, with
      each line in between representing a 0.5 increment in
      \textrm{[Fe/H]}; solid lines represent an age of 12.5 Gyr and
      the dashed lines represent 4 Gyr.  (A color version of this
      figure is available in the online journal.)  \label{agez2}} 
\end{figure}

Figure \ref{agez2} demonstrates the age independence and metallicity
separation of a color-color diagram. The plot includes isochrones of
two different ages, 12.5 and 4 Gyr, as solid and dashed lines
respectively, for a range of metallicities in steps of 0.5 dex.  The
solid and dashed lines closely follow each other throughout the
color-color diagram.  The bifurcation of the isochrones at cooler
temperatures (especially seen at higher metallicity with the dashed
and solid purple lines) is due to dwarf and giant stars colors showing
increased gravity sensitivity at higher metallicity. The gravity
sensitivity causes the metallicity-sensitive color of dwarfs to be
bluer than giants at the same value of the temperature-sensitive
color, however, for targets at a common distance, apparent magnitude
can be used to separate the evolutionary stage. Based upon the color
sensitivities shown in Figure \ref{dtdz}, we selected the more
metallicity-sensitive colors to use with the temperature-sensitive
color (F555W--F814W) in color-color diagrams.

\begin{figure}[ht]
  \epsfig{file=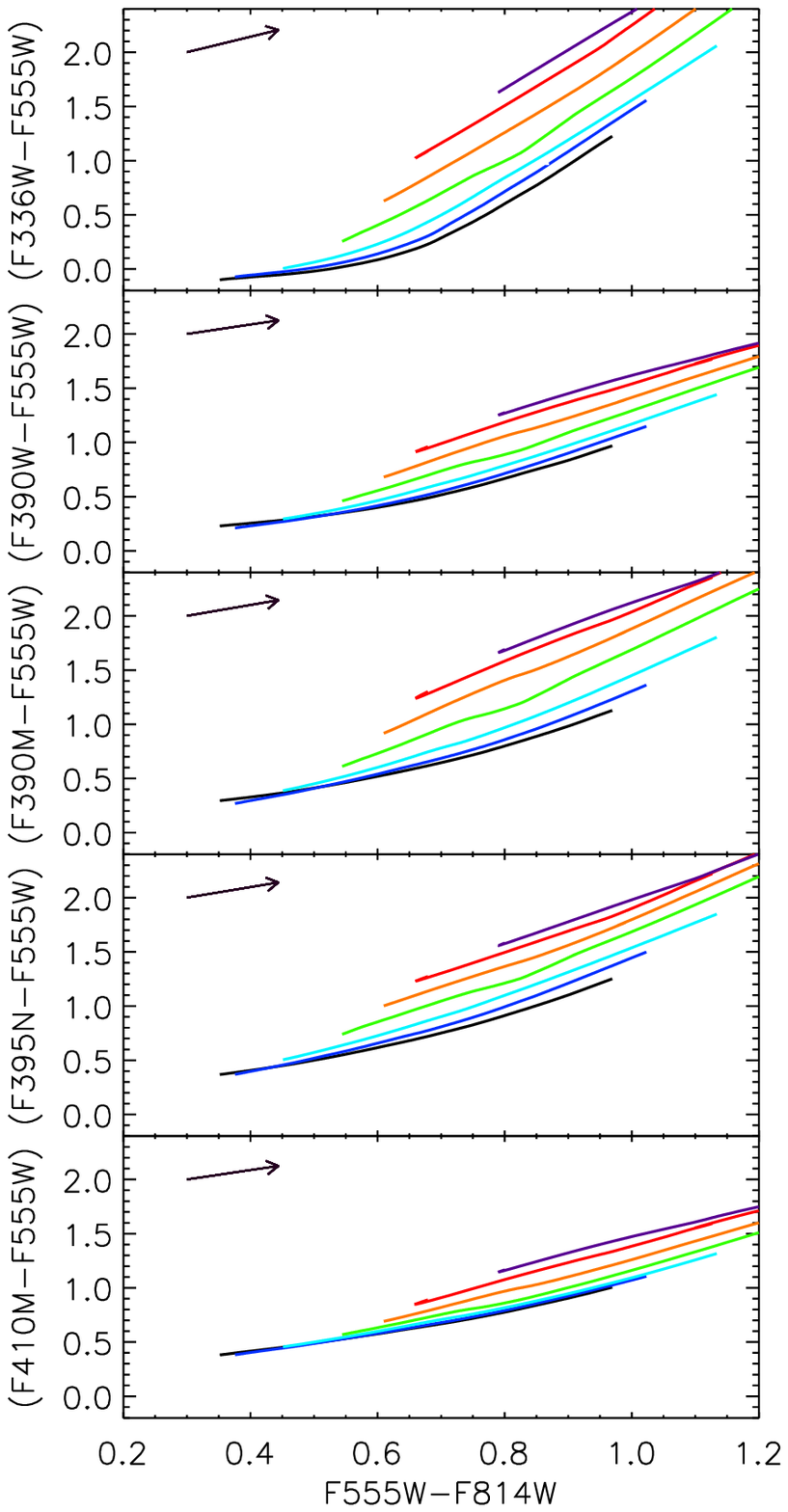,height=5.5in}
\caption{We show the metallicity sensitivity for different filter
  combinations using color-color plots of Dartmouth Isochrones above
  the MSTO. The lines span $-2.5 <$ \textrm{[Fe/H]} $< +0.5$, lowest
  (black) line being the most metal poor, the highest (purple) line
  the most metal rich.  Reddening vectors are indicated with arrows.
  (A color version of this figure is available in the online
  journal.)  \label{sensitivity}}  
\end{figure}

Figure \ref{sensitivity} shows giant branch isochrones in color-color
diagrams for five metallicity sensitive colors: F336W--F555W,
F390M--F555W, F390W--F555W, F395N--F555W and F410M--F555W. The
effectiveness of these color combinations to determine metallicity are
quantified in Table \ref{spans} by the color separation due to
metallicity predicted from stellar isochrones in color-color diagrams,
using units of dex of \textrm{[Fe/H]}/0.01 mag of color change. While all the
filter sets have similar sensitivity around solar metallicity ($+0.5
>$ \textrm{[Fe/H]} $> -0.5$) this breaks down at lower metallicities.

\begin{deluxetable*}{lcccccccc}[h]
\tablecaption{Color-Color diagram sensitivity: Metallicity per 0.01 magnitude color change}
\tablehead{
\multicolumn{1}{c}{}&
\multicolumn{1}{c}{}&
\multicolumn{5}{c}{ Sensitivities}&
\multicolumn{1}{c}{}&
\multicolumn{1}{c}{}\\
     \colhead{Color}  &
     \colhead{@F555W--}&
     \colhead{$-4.5 <$}&
     \colhead{$-3.5 <$}&
     \colhead{$-2.5 <$}&
     \colhead{$-1.5 <$}&
     \colhead{$-0.5 <$}&
     \colhead{normalized}& 
     \colhead{\textrm{[Fe/H]} /} \\					 
        \colhead{}  &
        \colhead{F814W= }&
        \colhead{\textrm{[Fe/H]}}&
        \colhead{\textrm{[Fe/H]}}&
        \colhead{\textrm{[Fe/H]}}&
        \colhead{\textrm{[Fe/H]}}&
        \colhead{\textrm{[Fe/H]}}&
        \colhead{$\sigma_{color}$}       &
        \colhead{(0.01 mag /}  \\
           \colhead{}  &
           \colhead{}  &
           \colhead{$< -3.5$}&
           \colhead{$< -2.5$}&
           \colhead{$< -1.5$}&
           \colhead{$< -0.5$}&
           \colhead{$<  0.5$}&
           \colhead{}        &
           \colhead{$\sigma_{color}$)} \\
\multicolumn{1}{c}{}&
\multicolumn{1}{c}{}&
\multicolumn{5}{c}{ (dex \textrm{[Fe/H]}/0.01 mag of color change) }&
\multicolumn{1}{c}{}&
\multicolumn{1}{c}{}}

\startdata
F336W--F555W &  0.9  & 0.427   & 0.143  &0.047  &0.023  &0.026   &  9.8  & 0.24 \\
             &  1.0  & 0.290   & 0.107  &0.045  &0.023  &0.024   &  9.8  & 0.23 \\
             &  1.1  & 0.236   & 0.087  &0.051  &0.023  &0.024   &  9.8  & 0.23 \\
\hline 				                       						                 
F390W--F555W &  0.9  & 0.943   & 0.287  &0.098  &0.041  &0.049   &  1.0  & 0.04 \\
             &  1.0  & 0.704   & 0.199  &0.090  &0.038  &0.046   &  1.0  & 0.04 \\
             &  1.1  & 0.541   & 0.149  &0.084  &0.038  &0.043   &  1.0  & 0.04 \\
\hline 				                      						                 
F390M--F555W &  0.9  & 0.667   & 0.206  &0.074  &0.027  &0.036   &  6.9  & 0.19 \\
             &  1.0  & 0.493   & 0.146  &0.062  &0.024  &0.035   &  6.9  & 0.16 \\
             &  1.1  & 0.370   & 0.110  &0.052  &0.023  &0.034   &  6.9  & 0.16 \\
\hline 				                     						                 
F395N--F555W &  0.9  & 0.515   & 0.151  &0.066  &0.036  &0.049   & 15.4  & 0.56 \\
             &  1.0  & 0.369   & 0.104  &0.057  &0.037  &0.047   & 15.4  & 0.56 \\
             &  1.1  & 0.277   & 0.079  &0.052  &0.038  &0.047   & 15.4  & 0.60 \\
\hline 				                     						                   
F410M--F555W &  0.9  & $>$ 1   & 0.820  &0.254  &0.090  &0.063   &  2.6  & 0.23 \\
             &  1.0  & $>$ 1   & 0.510  &0.288  &0.070  &0.051   &  2.6  & 0.16 \\
             &  1.1  & $>$ 1   & 0.364  &0.350  &0.063  &0.044   &  2.6  & 0.16 
\enddata
\tablecomments{Quantified color spans between isochrones 1 dex [Fe/H]
  apart in the color-color diagrams shown in Figure
  \ref{sensitivity}. All numbers are listed in dex of [Fe/H] /0.01 mag
  of color change. The  $\sigma_{color}$ is estimated using the online
  WFC3 ETC and normalized to $\sigma_{(F390W--F555W)}$. The last
  column reports the \textrm{[Fe/H]} dex /(0.01 mag/$\sigma$) for the
  metallicity range $-1.5 <$ \textrm{[Fe/H]} $<-0.5$. At extremely low
  metallicities some of the color changes are beyond typical
  photometric accuracy, i.e. greater than a dex of color / 0.01 mag of
  color change.  \label{spans} }   
\end{deluxetable*}

Weighting the metallicity sensitivity by the relative uncertainty
(last column of Table \ref{spans}) shows that F390W--F555W is the most
sensitive for a fixed exposure time.  Below \textrm{[Fe/H]} $<-1.5$ the color
separation for F395N--F555W, F390M--F555W and F336W--F555W are the
most sensitive to metallicity. Although the F395N--F555W color retains
sensitivity it also requires 2 to 3 times more exposure time to get
comparable accuracy, making it prohibitive to use.  At extremely low
metallicity the most sensitive color is F395N--F555W, with sensitivity
decreasing respectively for the colors:  F336W--F555W, F390M--F555W
and F390W--F555W. 
 
Based upon the metallicity sensitivity and photometric efficiency we
find that the most promising metallicity indicating filters are the
F390W, F390M and F336W filters. In the remainder of the paper we will
focus our analysis on these filters.

\subsection{Reddening Effects}

Reddening adds uncertainty to any photometric metallicity derivation,
especially when the uncertainty in reddening is large, or if there is
differential reddening throughout the field. In a color-color plot
uncorrected reddening will be confused with a change in metallicity if
the reddening vector is in the direction of the metallicity
separation.  For the majority of the color-color diagrams the 
reddening vectors are predominantly in the same direction as the
isochrones, mitigating the influence of reddening. However, for
(F336W--F555W, F555W--F814W) the reddening vector is $\sim$
40$^{\circ}$ shallower than the isochrones, which increases the
uncertainty in derived metallicities when there is differential
reddening or if the uncertainty in the reddening is large. Reddening 
vectors are indicated in Figure \ref{sensitivity} for E(B--V) = 0.1.
The vector is defined by the reddening of the colors along the X and Y
axes. For a given color, the reddening E(Filter$_1$--Filter$_2$) =
$A_{Filter1} -  A_{Filter2}$, where A$_{Filter1}$ = R$_{Filter1}$ E(B--V), 
and the R$_{Filter}$ values are listed in Table \ref{throughput}.   

To get an estimate of the effect of reddening on metallicities we
consider an uncertainty of $\sigma_{ E(B-V)}$ = 0.02, as might be
appropriate in a low reddening target. At shorter wavelengths, where
all of our metallicity sensitive filters are located, the uncertainty
in reddening will have an increased effect due to the UV extinction
law increasing as wavelength decreases.   In this case the
$\sigma_{E(B-V)}$ adds a systematic color uncertainty of $\sim$0.1 mag
to all of the metallicity sensitive colors.   However, the color
change from reddening will be along the reddening vector, and because 
the angles between the isochrones and reddening vectors vary (e.g. Figure
\ref{sensitivity}) and the colors have different sensitivity to
metallicity (e.g$.$ Table \ref{spans}), this leads to metallicity
uncertainties ranging from a tenth to a few tenths of a dex.  The
resulting systematic uncertainties in \textrm{[Fe/H]} are reported in Table
\ref{error}. As expected, larger uncertainties are found at lower
metallicities.

Historically, reddening-free indices have been proposed as a means to
correct colors for extinction.  A reddening-free index, Q, is defined
as: 
\begin{equation}
 Q=(m_1-m_2) - \frac{E(m_1 - m_2)}{E(m_2-m_3)}(m_2-m_3)
\end{equation}
where E$(m_1 - m_2)$ is the color excess for the color $(m_1 - m_2)$
\citep{1953ApJ...117..313J}.  However, making our metallicity
indicator reddening-free still leaves the problem that the temperature
indicator is affected by reddening. To completely remove reddening
effects from the color-color diagram, both indices need to be
reddening-free. \citet{2009AJ....137.3172B} suggest using two
reddening-free indices (5 total filters, including two in the IR) to
predict metallicity.  We looked into using these additional filters
and found that with our low reddening targets the additional filters
do not significantly add to the metallicity resolution.

\begin{deluxetable}{lccc}[Hb]
\tablefontsize{\tiny}
\tablecaption{Metallicity errors from Reddening}
\tablehead{
\multicolumn{1}{l}{}&
\multicolumn{3}{c}{Metallicity Errors}\\
     \colhead{Color}  &
     \colhead{-2.5 $<$}&
     \colhead{-1.5 $<$}&
     \colhead{-0.5 $<$} \\
       \colhead{}  &
       \colhead{\textrm{[Fe/H]}}&
       \colhead{\textrm{[Fe/H]}}&
       \colhead{\textrm{[Fe/H]}} \\
		 \colhead{}  &
		 \colhead{$<$ -1.5}&
		 \colhead{$<$ -0.5}&
		 \colhead{$<$  0.5} }
\startdata
F336W--F555W  & 0.30   & 0.14   & 0.15   \\
F390M--F555W  & 0.20   & 0.09   & 0.13   \\
F390W--F555W  & 0.23   & 0.11   & 0.12   \\
F395N--F555W  & 0.10   & 0.07   & 0.09   \\
F410M--F555W  & 0.33   & 0.12   & 0.08   
\enddata
\tablecomments{Metallicity uncertainties for each color is based on
  the metallicity sensitivity, the angle between the reddening vector and
  isochrone sequence, and $\sigma_{\rm{ E(B-V)}} = \pm$ 0.02.   \label{error}} 
\end{deluxetable}

\subsection{Abundance Ratio Variations}
While \textrm{[Fe/H]} is commonly used as a proxy for the overall stellar
metallicity, elements do not vary in lockstep from star to
star. Variations in abundance ratios can alter spectral features
within a bandpass, changing the stellar color and causing the overall
photometric metallicity to be misinterpreted if fixed abundance ratios
are assumed.   

\begin{figure}[ht]
  \epsfig{file=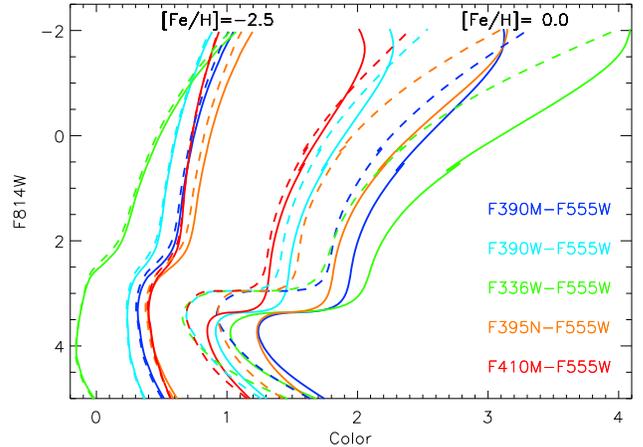,height=2.4in}
\caption{Color changes due to shifts in $\alpha$ at a low and high
  metallicity, for 5 metallicity-sensitive colors listed.  Solid and
  dashed lines represent [$\alpha$/Fe] = $+0.4$ and 0.0, at
  \textrm{[Fe/H]} = $-2.5$, and [$\alpha$/Fe] = $+0.2$ and $-0.2$ at
  \textrm{[Fe/H]} $= 0.0$. 	\\	(A color version of this figure is
  available in the online journal.)  \label{cmdafeh}}  
\end{figure}

A common abundance variation is $\alpha$ element enhancement
(e.g. \textrm{ Ne, Mg, Si, S, Ar,  Ca, and Ti}), often present at low
metallicity in globular clusters (see
\citet{2009ARA&A..47..371T,2010nuco.confE...8M} and references there
in).  The CMD of an $\alpha$-enhanced population resembles a CMD of
higher \textrm{[Fe/H]} with solar abundance ratios. Figure
\ref{cmdafeh} shows how $\alpha$ enhancements affect different colors
at low and high metallicity. Different combinations of \textrm{[Fe/H]}
and [$\alpha$/Fe] can give the same colors. 

Using Dartmouth isochrones, we compute the relative color change from
$\alpha$ abundances ($\Delta$ color/ $\Delta$ [$\alpha$/Fe]) and
metallicity ($\Delta$ color/ $\Delta$ \textrm{[Fe/H]}), to find the
relative sensitivity of various colors to \textrm{[Fe/H]} and $\alpha$
enhancement. The ratio of the two as a function of metallicity is
presented in Table \ref{da_dz}; unsurprisingly the colors with the
most $\alpha$ sensitivity are the the two colors (F395N--F555W and
F390M--F555W) whose bandpasses are dominated by the \textrm{Ca} H $\&$
K lines.  
  
Metallicities assigned assuming solar abundance ratios can be
considered as a metallicity indicator that is a combination of the
actual \textrm{[Fe/H]} and [$\alpha$/Fe], following the relation:

\begin{equation}
\textrm{[m/H]}_{phot}=\textrm{[Fe/H]}+ \left( \frac{\Delta \textrm{[Fe/H]} }{
  \Delta[\alpha/\textrm{Fe}]} \right) [\alpha/\textrm{Fe}]
\end{equation}
where the amount of relative color change due to $\alpha$ enhancements
(listed in Table \ref{da_dz}) can be used to get the relation.

The differing sensitivity of ($\Delta$\textrm{[Fe/H]}/$\Delta$[$\alpha$/Fe]) 
in different colors implies that $\alpha$ abundance can be separated with 
photometry using similar techniques described in Section 2.3. However,
additional colors with differing sensitivities  would be necessary to
isolate the $\alpha$ abundance, such as F395N--F555W and F410M--F555W
or F390M--F555W and F390W--F555W. 

We should note that the majority of globular clusters show large
star-to-star variations in light elements (Li to Si) (for a review see
Gratton et al$.$ 2012, and references therein).
\nocite{2012A&ARv..20...50G} Additionally there are cluster-to-cluster
abundance variations \citep{2009A&A...508..695C}, which also have the
potential to affect the stellar colors.  \citet{2012ApJ...755...15V}
found that abundance enhancements of Mg, Si, and to some extent Ca,
will change the color and thus location of the giant branch, while
variations in oxygen will affect the height of the subgiant branch and
the luminosity of the MS turnoff.  Of particular interest in this
study, C and N enhancements can affect the stellar colors measured
with F390M and F390W, since a few CN and CH bands fall in the same
spectral region covered by these filter bandpasses; see Section 2.1
for details.  Since we cannot explicitily measure any of these
abundance variation in this study due to the limited resolution of
photometric metallicities, any such variations could affect the
stellar colors and our inferred metallicities.

\begin{deluxetable}{lccc}[Hb]
\tablecaption{Relative sensitivity of \textrm{[Fe/H]} to [$\alpha$/Fe] }
\tablehead{
\multicolumn{1}{l}{}&
\multicolumn{3}{c}{$\Delta$\textrm{[Fe/H]} / $\Delta$ [$\alpha$/Fe] }\\
     \colhead{Color}  &
     \colhead{$-2.5 <$ }&
     \colhead{$-1.5 <$ }&
     \colhead{$-0.5 <$ } \\
       \colhead{}  &
       \colhead{ \textrm{[Fe/H]} }&
       \colhead{ \textrm{[Fe/H]} }&
       \colhead{ \textrm{[Fe/H]} } \\
		 \colhead{}  &
		 \colhead{ $< -1.5$}&
		 \colhead{ $< -0.5$}&
		 \colhead{ $<  0.5$} }
\startdata
F336W--F555W  &  0.27  &  0.39  &  0.64  \\
F390W--F555W  &  0.22  &  0.19  &  0.09  \\
F390M--F555W  &  0.65  &  0.37  &  0.34  \\
F395N--F555W  &  0.88  &  0.84  &  0.79  \\
F410M--F555W  &  0.09  &  0.17  & -0.09
\enddata
\tablecomments{The relative sensitivity of metallicity compared to
  $\alpha$ enhancement   \label{da_dz}}
\end{deluxetable}

\section{Testing the method: Observations of calibration clusters}
WFC3 images were obtained of five clusters during cycle 17 as part of
the calibration program 11729, (PI Holtzman).  Additional images of
the five clusters come from the calibration portion of the program
11664 (PI Brown). \nocite{2008hst..prop11664B} The five clusters were
chosen because they are well studied and span a range of \textrm{[Fe/H]}; 5 of
the 6 clusters are discussed by \citet{2005AJ....130.1693B} for ACS
calibration. \nocite{2008hst..prop11729H}

\begin{deluxetable*}{lccccc}[H]
\tablefontsize{\tiny}
\tablecaption{Exposure Times}
\tablehead{
 \multicolumn{1}{c}{WFC3}&
 \multicolumn{5}{c}{Exposure Time (sec)}\\
     \colhead{Filter}  &
     \colhead{M 92 }&
     \colhead{NGC 6752 }&
     \colhead{NGC 104 }&
     \colhead{NGC 5927 }&
     \colhead{NGC 6791 }      }
\startdata
F336W   &850, 30    &1000, 30     &1160, 30     &950, 30     &800, 30     \\
F390M   &1400, 50   &1400, 50     &1400, 50     &1400, 50    &1400, 50    \\
F390W   &2290, 10   &2460, 10     &2596, 10     &2376, 10    &2250, 10	  \\
F395N   &1930, 90   &2100, 90     &2240, 90     &1015, 90    &1850, 90    \\
F410M   &1530, 40   &1600, 40     &1600, 40     &1600, 40    &1490, 40    \\
F467M   &700, 40    &800, 40      &900, 40      &730, 40     &700, 40     \\
F547M   &445    &445      &445      &445     &445     \\ 
F555W   &1360.5 &1360.5   &1360.48  &1381    &1360.5  \\
F814W   & 860.5 &1020.5   &1160.48  &961     &810.5   \\
F160W	&1245, 8.3 &1245, 12.5   &1245, 16.7    &1245, 12.5  &1245, 4.2   \\ 
F110W	&1245, 8.3  &1245, 12.5   &1245, 16.7   &1245, 12.5   &1245, 4.2  
\enddata
\label{exposure}
\end{deluxetable*}

\subsection{The Observations}
We present data for four globular clusters, M92 (NGC 6341), NGC 6752, NGC 104
(47 Tuc), NGC 5927, and one open cluster NGC 6791.  
Exposure times are listed in Table \ref{exposure}, and include short
and long exposures to increase the dynamic range of the photometry.

\subsection{Photometry}

Reduced images were obtained from the HST archive using its
on-the-fly processing. A deep image was created by summing all of the
exposures in F467M, and stars were identified on this image. For each
individual frame, a pixel area correction was applied to account for
the modification in fluxes by flat fielding, and an astrometric
solution was derived relative to the reference frame. Aperture
photometry was done on all of the stars using aperture radii of 0.12,
0.25, 0.375, and 0.5 arcsec. Using this initial aperture photometry
and model PSFs calculated using TinyTim, photometry was redone on each
individual star after subtracting off all neighbors; this process was
iterated twice.

\subsection{Cluster parameters}
To adopt isochrones for the clusters we search the literature
for well established cluster parameters. The distance modulus and the
reddening are needed to get the absolute magnitudes of the cluster
stars.  Additionally we use the age, metallicity, and $\alpha$
enhancement to select isochrones for our study.  We have compiled a
sampling of the reported cluster parameters, and list the literature
values adopted in this study in Table \ref{obsdat}.

{\bf M92} is the only very metal-poor cluster in our sample.
Spectroscopic metallicity measurements include those by
\citet{1984ApJS...55...45Z}, \textrm{[Fe/H]} $= -2.24 \pm 0.08$,  
\citet{1997A&AS..121...95C}, \textrm{[Fe/H]} $= -2.16 \pm 0.02$, and
\citet{2003PASP..115..143K},$\langle$ \textrm{[Fe/H]$_{II}$}$\rangle =
-2.38 \pm 0.02$ and $\langle$ \textrm{[Fe/H]$_{I}$}$\rangle =
-2.50 \pm 0.12$..  We average the spectroscopic metallicities, and
adopt \textrm{[Fe/H]} $= -2.30$. Isochrone studies by
\citet{2003AJ....126..778V} and \citet{2010PASP..122..991D} both found
$\alpha$ enhancements of [$\alpha$/Fe] $= +0.3$, however our isochrone
grid is in steps of 0.2, so we adopt [$\alpha$/Fe]$=0.4$. We adopt the
age, and reddening found by \citet{2003AJ....126..778V},  13.5 Gyr and
E(B--V) $= 0.023$.  After adopting these isochrone parameters we
adjust the distance modulus to be fully consistent with our isochrone
models, by assuming (m--M)$_{\rm{V}} = 14.60$, which is consistent
with the values used by \citet{2003AJ....126..778V}.

{\bf NGC 6752}. \citet{2005A&A...440..901G} measured spectra of seven
stars for an overall metallicity \textrm{[Fe/H]} $= -1.48 \pm0.02$,
and [$\alpha$/Fe] $=+0.27$ $\pm$0.01.  \citet{2009A&A...508..695C}
measured \textrm{[Fe/H]} $=-1.555 \pm 0.07$ from high
resolution spectra of 14 stars, they also measured multiple $\alpha$
elments for an average [$\alpha$/Fe] $= +0.35$.  In a previous study,
\citet{2003A&A...408..529G} found \textrm{[Fe/H]} $= -1.43\pm
0.04$. Other spectroscopic metallicity measurements include those by
\citet{1984ApJS...55...45Z}, \textrm{[Fe/H]} $= -1.54 \pm 0.09$,
\citet{1997A&AS..121...95C}, \textrm{[Fe/H]} $= -1.42 \pm 0.08$,
\citet{2003PASP..115..143K}, $\langle$ \textrm{[Fe/H]$_{II}$}$\rangle
= -1.50 \pm 0.02$.  We average the spectroscopic \textrm{[Fe/H]} and
adopt a \textrm{[Fe/H]} $= -1.45$, which is slightly lower than the
average but is consistent with the larger [$\alpha$/Fe] adopted to
fit our isochrone grid, [$\alpha$/Fe]$=0.4$. 

We assume a reddening of  E(B--V) $= 0.046 \pm 0.005$ as measured by
\citet{2005A&A...440..901G} from 118 stars. We adopt an age of 12.5
Gyr, reported by \citet{2000ApJS..129..315V}, and also by
\citet{2009ApJ...694.1498M} in a study of comparative ages of globular
clusters.  From these adopted isochrone prameters we adjust our
distance modulus to (m--M)$_{\rm{V}} = 13.20$, which is consistent
with the distance to NGC 6752 measured by \citet{1996ApJ...465L..23R}
using the white dwarf cooling sequence; (m--M)$_{\rm{V}}$= 13.17.

{\bf 47 Tuc}.  Spectroscopic measurement include those by
\citet{1984ApJS...55...45Z}, \textrm{[Fe/H]} $= -0.71 \pm 0.08$,
\citet{1997A&AS..121...95C} who found \textrm{[Fe/H]} $=-0.70 \pm
0.07$, \citet{2003PASP..115..143K}, $\langle$ \textrm{[Fe/H]$_{II}$}
$\rangle = -0.70 \pm 0.05$, \citet{2006ApJ...649..248W} found
\textrm{[Fe/H]} $=-0.60 \pm 0.2$ based on seven RGB stars, and
\citet{2009A&A...508..695C} measured \textrm{[Fe/H]} $=-0.74 \pm 0.02$
from the medium resolution spectra of 147 stars and \textrm{[Fe/H]}
$=-0.77 \pm 0.03$  from the high resolution spectra of 11 stars, and
measured an [$\alpha$/Fe]$=0.44 $.  We average the \textrm{[Fe/H]}
adopting values of \textrm{[Fe/H]} $= -0.70$ and [$\alpha$/Fe] = 0.4.
  
We use the E(B--V) $= 0.024$ reported by \citet{2003A&A...408..529G}.
\citet{2003AJ....126..778V} found an age of 12 Gyr,
\citet{2001ApJ...553..733Z} derived an age of 12.9 Gyr using diffusive
models, which we average for an adopted age of 12.5 Gyr. 
Based upon the adopted isochrone parameters we adjust the distance
modulus to (m--M)$_{\rm{V}}=13.20$ which is smaller than the values
measured by \citet{2012AJ....143...50W} and
\citet{2001ApJ...553..733Z} who both  measured the distance modulus
(13.36 and 13.27, respectively) to 47 Tuc using white dwarf cooling models.

\begin{figure*}[ht]
  \epsfig{file=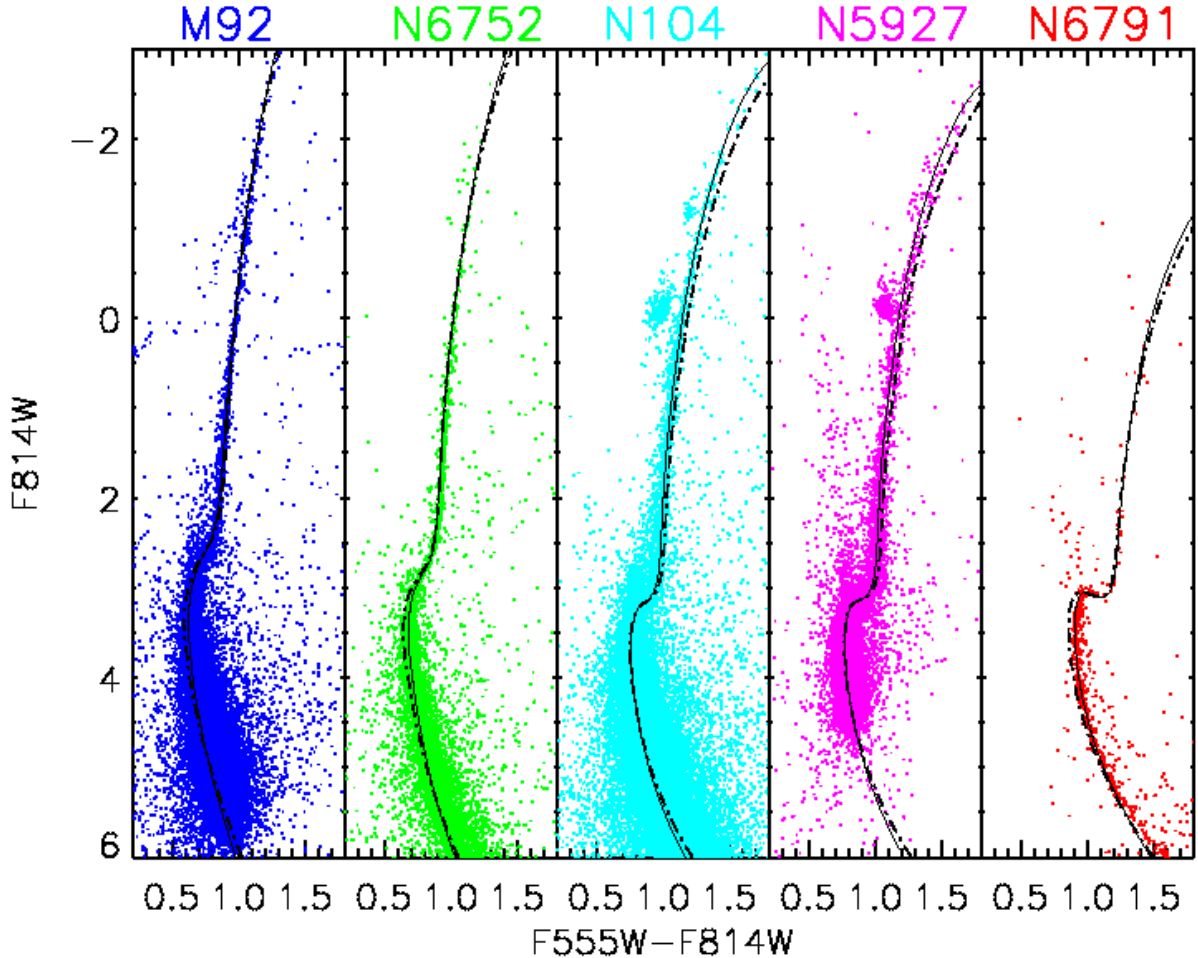, height=5.3in }
  \caption{CMD of (F555W--F814W, F814W), and Dartmouth isochrones of
  literature value (dashed lines) as listed in Table \ref{obsdat}. 
  Empirically corrected isochrones shown as solid lines,
  corrections listed in Table 8. 	(A color version of this
  figure is available in the online journal.)  \label{CMDs} }
\end{figure*}

{\bf NGC 5927} is more complicated to study photometrically due to the
high differential reddening within the cluster ( $\Delta$ E(V-I)=0.27
measured by \citet{1999A&A...347..455H} and $\delta$(E(B--V)=0.169
measured by \nocite{2013MmSAI..84..187B} Bonatto et al. 2013).
\citet{1984ApJS...55...45Z} found \textrm{[Fe/H]}  $=-0.30\pm0.09$;  
\citet{1988AJ.....96...92A} measured \textrm{[Fe/H]} $= -0.31$, while
\citet{2003PASP..115..143K} found \textrm{[Fe/H]} $ = -0.67$. We adopt
a \textrm{[Fe/H]} of $-0.40$ and [$\alpha$/Fe] of +0.2.  

\citet{1996AJ....112.1487H} found the reddening to be E(B--V)= 0.45.  
\citet{2005AJ....130..116D} did a study on the relative ages of GC's
and found the age to be $\approx$ 10 Gyr, while the relative GC age 
study by \citet{2009ApJ...694.1498M} determined an age of $\sim$ 11.3
Gyr,  we assume an age of 11.5 Gyr. Based upon the adopted isochrone
parameters we adjust the distance modulus to (m--M)$_{\rm{V}}=15.78$.

{\bf NGC 6791}, unlike the rest of our calibrating clusters, is an open
cluster (although recent work by \citet{2012arXiv1207.3328G} suggests
that NGC 6791 might be the remains of a stripped globular cluster).
\citet{2008A&A...492..171G} determined a metallicity of \textrm{[Fe/H]} $=
+0.37$. \citet{2006ApJ...646..499O} used infrared spectroscopy of 6
stars to measure a \textrm{[Fe/H]} = $+0.35 \pm$0.02 and solar $\alpha$
abundance ratio. \citet{1999AJ....117.1360C} found a \textrm{[Fe/H]}
$= +0.4$. We average the spectroscopic metallicities and round to the
nearest isochrone grid spacing and adopt \textrm{[Fe/H]} $=+0.4$ and
[$\alpha$/Fe]=0.0.  

We assume the reddening found by \citet{1999AJ....117.1360C}, E(B--V)
= 0.10$^{+0.03}_{-0.02}$.  We use the age found by both
\citet{1999AJ....117.1360C} and \citet{2006ApJ...643.1151C} of 8 Gyr.
\citet{2012A&A...543A.106B} found an age of $\sim$8.3 Gyr, and 
\citet{2013EPJWC..4305003G} also reported a consistent age of 8 Gyr
between the white dwarf cooling sequence and the MSTO age.  From this
we find a distance modulus of (m--M)$_{\rm{V}}=13.45$, which is
consisten with \citet{2008A&A...492..171G} who used an eclipsing
binary to determine (m--M)$_{\rm{V}}$ = 13.46$\pm$ 0.1, and with
\citet{1999AJ....117.1360C}, who found (m--M)$_{\rm{v}}
=13.45^{+0.03}_{-0.12}$.   

As previously stated the adopted \textrm{[Fe/H]} and [$\alpha$/Fe]
are not perfectly matched to the literature values but are rounded to
the spacing within the isochrone grid (in steps of 0.05 of
\textrm{[Fe/H]} and 0.2 of [$\alpha$/Fe]).  When adjusting the
metallicity to our grid we note that the \textrm{[Fe/H]} adjustment
for a change in [$\alpha$/Fe] varied for each color according to
equation (2) and Table \ref{da_dz}; therefore we made an average
metallicity adjustment that best fit all colors simultaneously. 

\begin{deluxetable*}{llrlllrr}[Ht]
\tabletypesize{\scriptsize}
\tablewidth{0pt}
\tablecaption{Cluster data}
\tablehead{
  \colhead{Cluster} &
  \colhead{RA}      &
  \colhead{Dec}     &
  \colhead{E(B--V)}  &
  \colhead{(m--M)$_{\rm{v}}$}   &
  \colhead{\textrm{[Fe/H]}\tablenotemark{a}  }   &
  \colhead{[$\alpha$/Fe]\tablenotemark{a}} &
  \colhead{Age}
}
\startdata
M 92     &17 17 07.05  &+43 07 58.2    &0.023  &14.60  &--2.50, --2.14 (--2.30)   &+0.30 (+0.4)  &13.5\\ 
NGC 6752 &19 10 54.86  &--59 59 11.2   &0.046  &13.20  &--1.54, --1.43 (--1.45)   &+0.27 (+0.4) &12.5\\
NGC 104  &00 24 15.26  &--72 05 47.9   &0.024  &13.20  &--0.77, --0.60 (--0.70)    &+0.44 (+0.4)  &12.5 \\
NGC 5927 &15 28 00.20  &--50 40 26.2   &0.45   &15.78  &--0.67, --0.30 (--0.40)   &. . . (+0.2)  &11.5\\ 
NGC 6791 &19 20 53.00  &+37 46 30.0    &0.10   &13.45  &+0.35,   0.40 ($+0.40$)   & 0.0 ( 0.0)   &8.0 
\enddata
\tablenotetext{a}{The range of spectroscopic metallicities found in
  the literature, followed (in parenthesis) by the isochrone
  metallicities adopted.}  
 \tablecomments{The cluster parameters adopted for this study including
  the reddening, metallicity, and ages from peer reviewed sources
  discussed and cited in section 3.3.  The distance moduli adopted for
  this study were adjusted to self consistently fit the MSTO and
  subgiant branchs for each cluster to the adopted isochrones. See
  Section 3.3 for more details.  	\label{obsdat}}   

\end{deluxetable*}

\begin{figure*}[ht]
  \epsfig{file=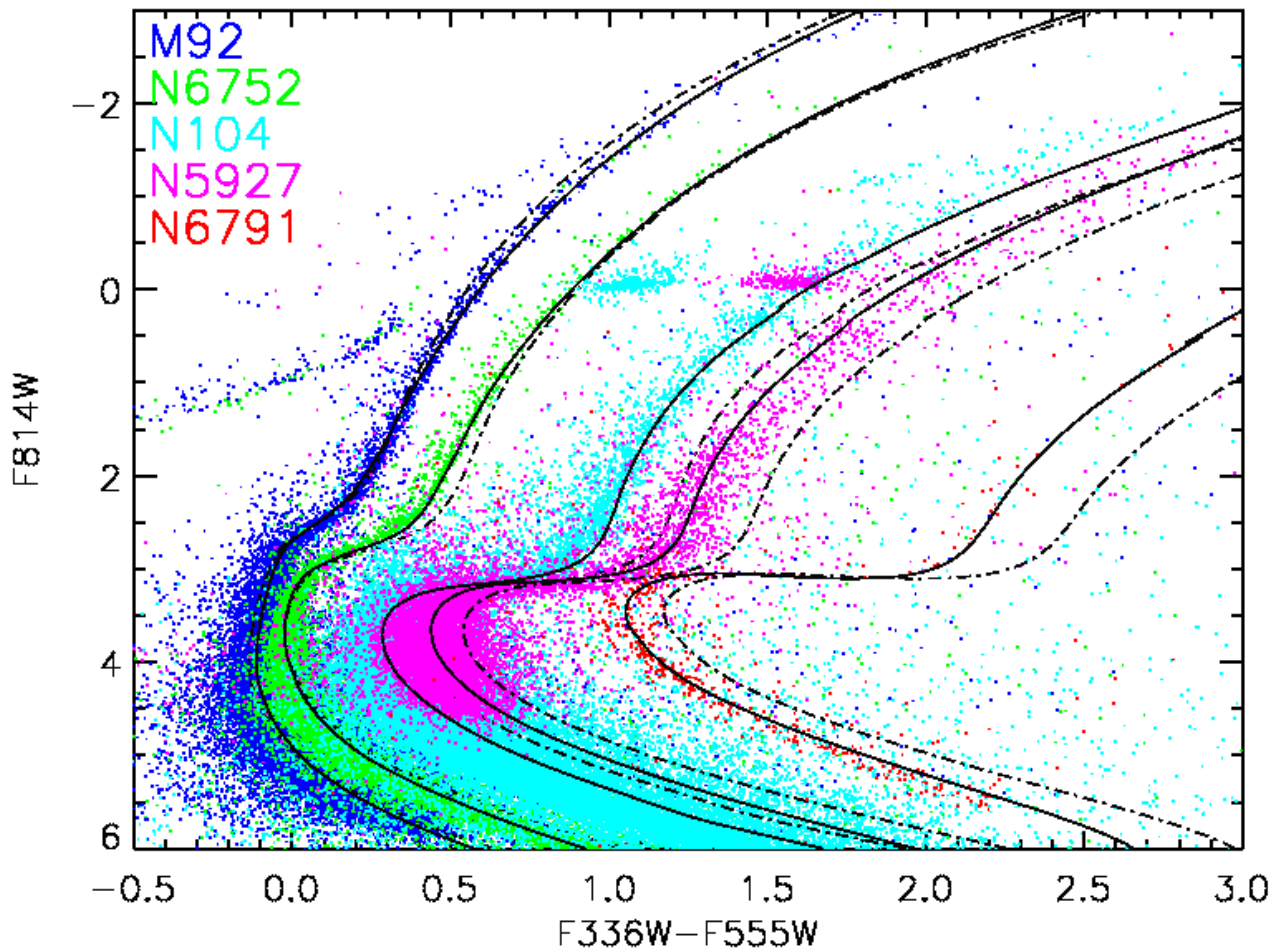, height=5.3in }
  \caption{CMD of (F336W--F555W, F814W) and Dartmouth isochrones of
  literature value (dashed lines) as listed in Table \ref{obsdat}. 
  Empirically corrected isochrones shown as solid lines,
  corrections listed in Table 8.  (A color version of this
  figure is available in the online journal.) \label{uCMD}}  
\end{figure*}

\subsection{Absolute Magnitudes}
From the reported reddening, and distance modulus we calculate the
Absolute magnitudes for each filter.  All magnitudes reported are in
the Vegamag system with zeropoints taken from the WFC3 handbook. The
absolute magnitudes for each filter were calculated using:
M$_{(filter)}=$ m$_{observed}$ - ((m--M)$_{\rm{v}}$ - A$_{\rm{v}}$) -
A$_{filter}$.  The distance modulus, (m--M)$_{\rm{v}}$, as
reported in Table \ref{obsdat}, was corrected for the V band
extinction,  $A_{\rm{v}}$ = 3.1E(B--V), and addtionally 
corrected for the extinction within the given filter A$_{(filter)}$ =
R$_{(filter)}$E(B--V). The ratio of total to selective
extinction, R$_{(filter)}$ as listed in Table \ref{throughput}, was
calculated by taking the integral over the stellar atmosphere SED
convolved with the filter transmission curve, divided by the integral
over the same atmosphere and filter convolved with the galactic
extinction curve. We used an atmosphere with \textrm{[Fe/H]} $=-1.5$,
temperature of 5000, and log g = 2.0.

\begin{figure*}[ht]
 \epsfig{file=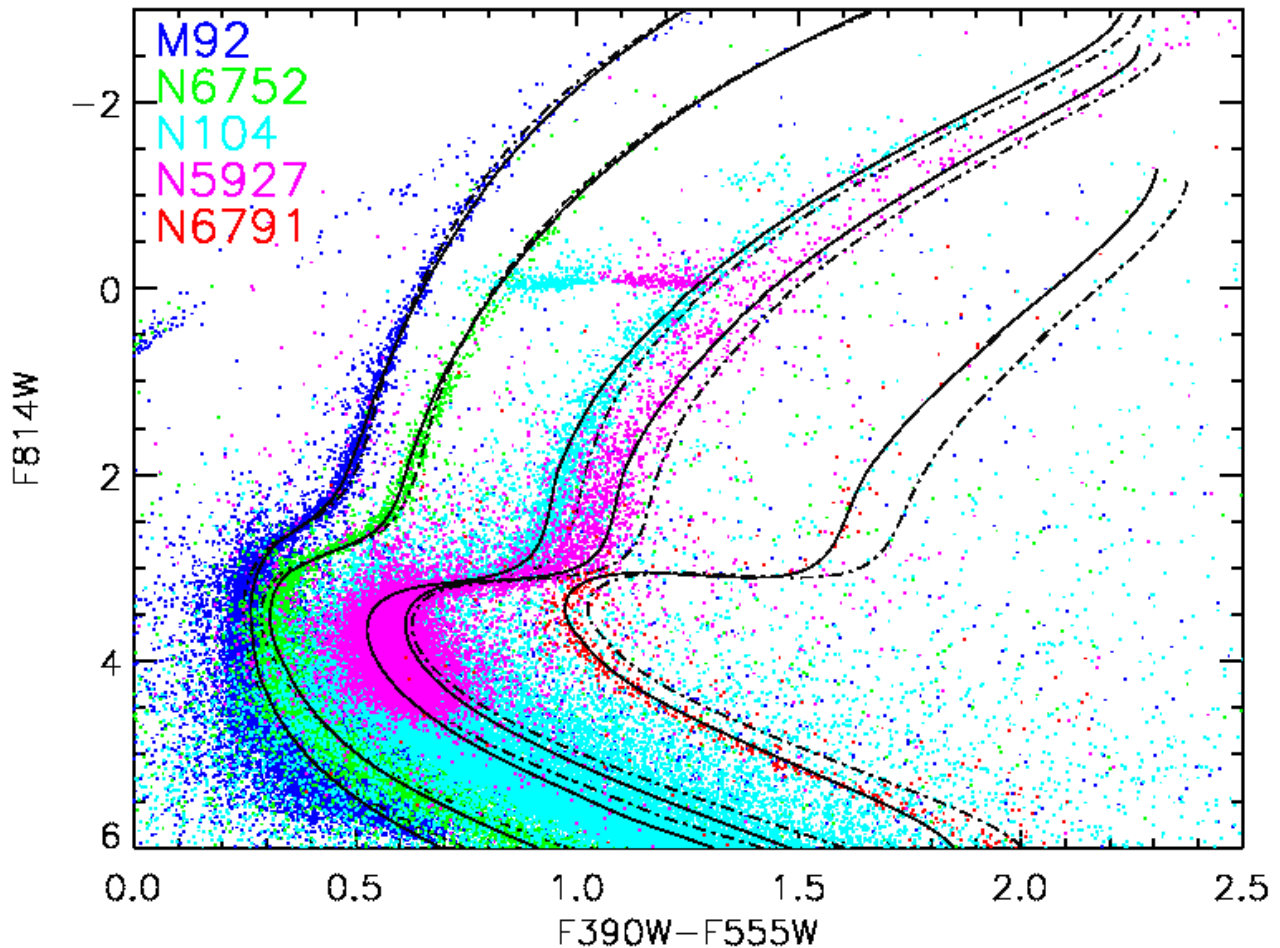, height=5.3in }
 \caption{CMD of F390W--F555W vs. F814W,and Dartmouth isochrones of
  literature value (dashed lines) as listed in Table \ref{obsdat}. 
  Empirically corrected isochrones shown as solid lines,
  corrections listed in Table 8. 	(A color version of this
  figure is available in the online journal.)  \label{wideCMD}} 
\end{figure*}

\section{Cluster Isochrone comparison}

\begin{figure*}[ht]
 \epsfig{file=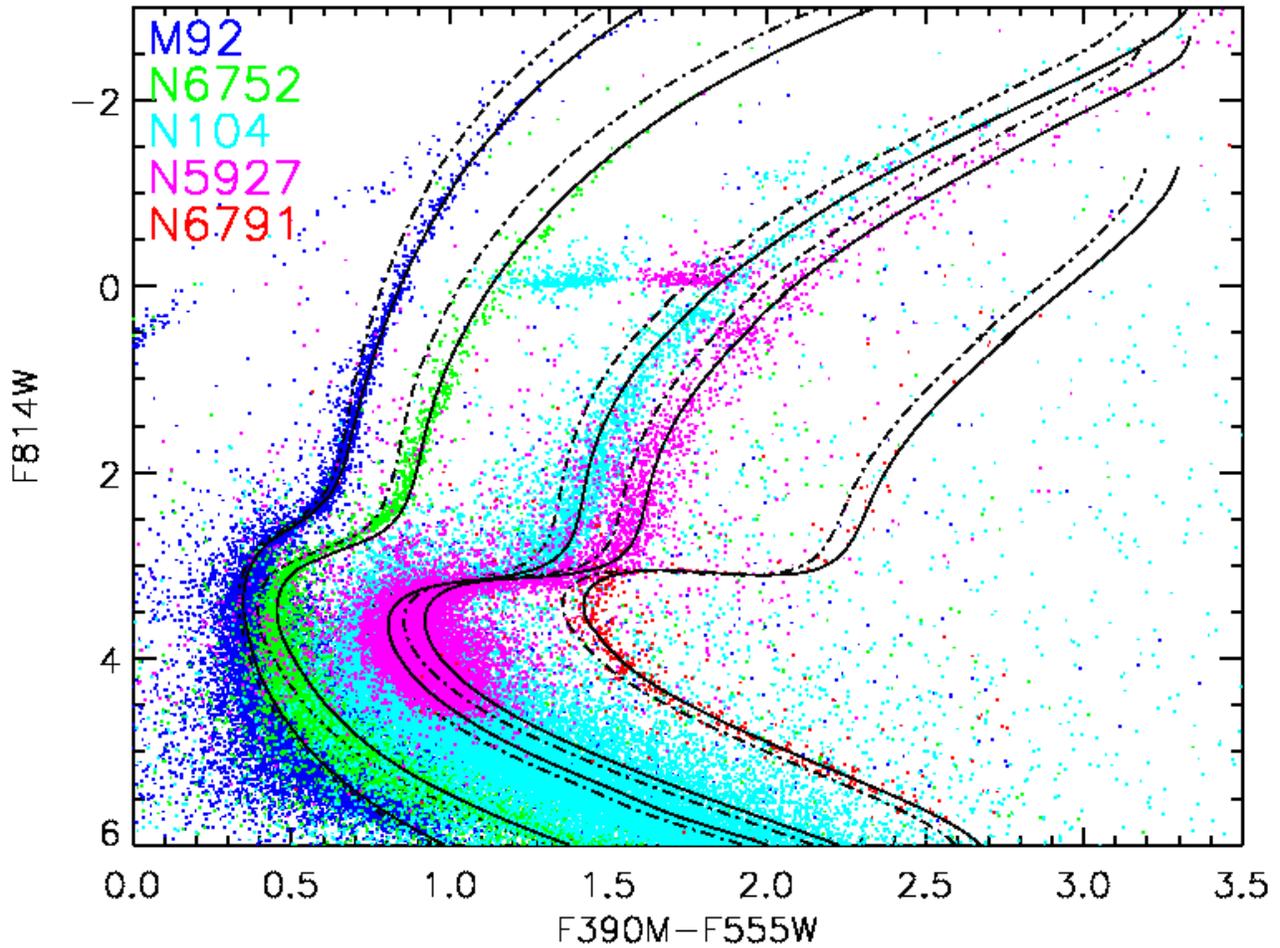,height=5.3in }
\caption{CMD of (F390M--F555W,  F814W), and Dartmouth isochrones of
  literature value (dashed lines) as listed in Table \ref{obsdat}. 
  Empirically corrected isochrones shown as solid lines,
  corrections listed in Table 8.  	(A color version of this
  figure is available in the online journal.) \label{medCMD}}
\end{figure*}

It is well known that theoretical isochrone models do not perfectly
match observed colors due to a combination of uncertainties from
stellar evolution models, model atmospheres, and instrumental
systematics.  When comparing the isochrone of the mean spectroscopic
value to the CMDs the models often did not match the shape of the
cluster ridgelines.  However, it should be noted that uncertainties in 
distance, age, reddening and composition complicate the fitting
process.   The cluster CMDs are shown in Figures \ref{CMDs} -
\ref{medCMD}, where dashed lines represent the literature valued
isochrones adopted for each cluster as reported in Table \ref{obsdat}.

In order to improve the metallicity determinations we derived an
empirical correction to the isochrones using the five cluster CMDs as
benchmarks.  For small steps in magnitude along the giant branch, we
found the distance between the theoretical isochrone colors and the
cluster ridgeline, then squared and summed the distances.  Using the
canonical metallicities we found the empirical corrections with the
smallest summed distance between the corrected isochrone and the
cluster. These corrections lead to consistent cluster metallicities
across all CMDs. All isochrone corrections were based on color,
gravity and metallicity adjustments. The various filter combinations
require different functional forms of the generic equation:

\begin{eqnarray}
{\rm(color)}_{\text{corrected}}=(\alpha +\beta\text{\textrm{[Fe/H]}}){\rm(color)}_{\rm{obs}} \\
+ \gamma(\text{\textrm{[Fe/H]}}+\delta)^2 +\epsilon 
\nonumber \\ + ( \zeta +\eta \textrm{[Fe/H]})(\theta - \text{log g}) \nonumber
 \end{eqnarray}

Equation (3) was modifed by inspection of the various CMDs. 
The (F555W--F814W, F814W) CMD in Figure \ref{CMDs} shows the
uncorrected literature valued isochrones falling redward of the
cluster ridgelines to varying degrees depending on the magnitude and
metallicity of the isochrone.  The most metal-poor cluster was
especially discrepant, with larger deviations along the lower giant
branch than the upper.  For this CMD the first order color correction
includes a dependence on metallicity, which allows for larger
corrections at lower metallicity. The first order color dependent term
shifts and changes the shape of the isochrone, while the second order
metallicity dependent term adds the same shift to a given isochrone
without changing the shape.  The additional gravity dependent
term corrects the bend of the giant branch (where log g $<$ 3.4), with
more bend required as metallicity decreases.

In Figure \ref{uCMD} the (F336W--F555W, F814W) CMD shows the
uncorrected isochrones redward of the cluster ridgelines, except for
M92 where the isochrone along the giant branch falls blueward. This
CMDs shows the largest difference between the clusters and the
models.  The bend of the isochrone along the giant branch (where log g
$<$ 3.4) is corrected with a gravity and metallicity dependent term,
with increasing tilt as metallicity decreases.  In this CMD the
corrected isochrone for 47 Tuc (NGC 104) falls  $\sim$0.03 mag redward
of the cluster ridgeline, while all of the other clusters are within
$\sim$0.01 mag of their ridgelines.  We believe this slight offset to be an
artifact of rounding the average spectroscopic metallicity to fit the
grid spacing.  This color offset is smaller than the color difference
between a isochrones of [Fe/H] $=-0.70$ and$ -0.75$, therefore we do
not force and overcorrection which would potentially lead to distorted
MDFs.

In Figure \ref{wideCMD} the uncorrected isochrones in the
(F390W--F555W, F814W) CMDs are the least discrepant of the CMDs we
examined. However a few corrections are still needed to improve the
fits, including a metallicity dependent gravity term, with smaller
corrections required for lower metallicities.

\begin{deluxetable}{lllll}[hb]
\tabletypesize{\scriptsize}
\tablenum{9}
\tablecaption{Cluster MDF Parameters from CMD fitting     }
\tablehead{
  \colhead{CMD} &
  \colhead{Cluster} &
  \colhead{Peak}      &
  \colhead{$\sigma$ }     &
  \colhead{$\#$ of }  \\
  \colhead{} &
  \colhead{} &
  \colhead{\textrm{[Fe/H]}} &
  \colhead{} &
  \colhead{stars}  }
\startdata

(F336W--F555W, F814W)          &      M92 &  -2.28  & 0.114  &     937   \\  
. . . . . . . . . . . . . . . .&    N6752 &  -1.45  & 0.073  &     481   \\  
. . . . . . . . . . . . . . . .&     N104 &  -0.74  & 0.059  &    2096   \\  
. . . . . . . . . . . . . . . .&    N5927 &  -0.37  & 0.067  &    1518   \\  
. . . . . . . . . . . . . . . .&    N6791 &   0.39  & 0.079  &     320   \\  
\hline							                                           
(F390M--F555W, F814W)          &      M92 &  -2.32  & 0.089  &     899   \\ 
. . . . . . . . . . . . . . . .&    N6752 &  -1.43  & 0.087  &     468   \\ 
. . . . . . . . . . . . . . . .&     N104 &  -0.68  & 0.066  &    2071   \\ 
. . . . . . . . . . . . . . . .&    N5927 &  -0.42  & 0.108  &    1504   \\ 
. . . . . . . . . . . . . . . .&    N6791 &   0.40  & 0.098  &     311   \\ 
\hline							                                           
(F390W--F555W, F814W)          &      M92 &  -2.33  & 0.092  &     906   \\ 
. . . . . . . . . . . . . . . .&    N6752 &  -1.39  & 0.060  &     462   \\ 
. . . . . . . . . . . . . . . .&     N104 &  -0.66  & 0.060  &    2067   \\ 
. . . . . . . . . . . . . . . .&    N5927 &  -0.46  & 0.124  &    1443   \\ 
. . . . . . . . . . . . . . . .&    N6791 &   0.39  & 0.082  &     297   
\enddata
\tablecomments{\label{mdfcmd}}
\end{deluxetable}

The (F390M--F555W, F814W) CMDs presented in Figure \ref{medCMD} shows
the uncorrected isochrones blueward of the clusters. Additionally the
isochrone curve along the giant branch is too steep to match the shape
of the clusters. We applied a first order color term that has an
additional metallicity dependence, making the color coefficient
greater for lower metallicities.

\begin{figure*}[ht]
\centering 
 \epsfig{file=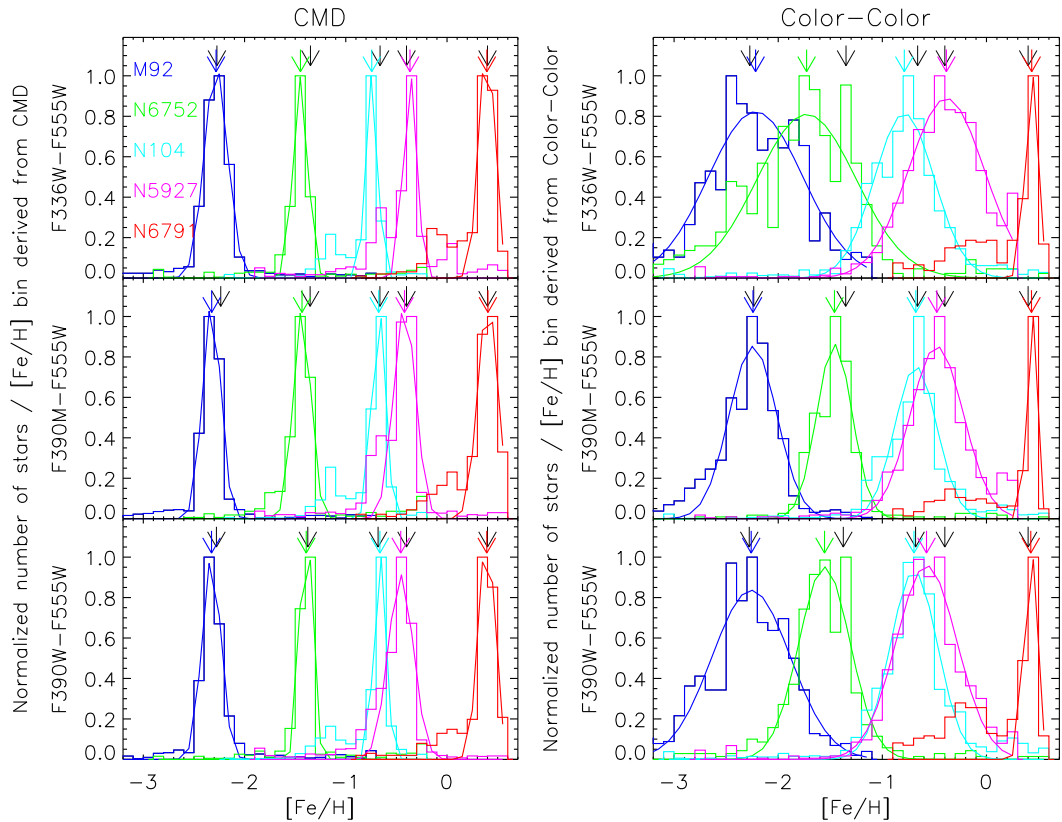,height=4.5in }
  \caption{The MDFs for M92, NGC 6752, NGC 104, NGC 5927 and NGC 6791
    (from left to right in each panel) as measured from CMDs and
    color-color diagrams. The left panels show the normalized
    distributions of metallicities measured by matching corrected
    isochrones to stellar colors from CMDs of F336W-F555W,
    F390M--F555W and F390W--F555W vs. F814W respectively. The right
    panels show the metallicities measured using corrected isochrones
    in color-color diagrams of the same colors vs. F555W--F814W. 
	Colored arrows indicate the mean of the distribution, black arrows
    indicate the metallicities from literature listed in Table
    \ref{obsdat} with the \textrm{[Fe/H]} adjusted for the adopted
    [$\alpha$/Fe] compared to the literature values, adjustments were
    small $\lesssim$ 0.05 dex. 	(A color version of this figure is
    available in the online journal.)  \label{plotmdf}}  
\end{figure*}

The isochrone corrections for all CMDs are listed in Table 8. Finding
a single correction to align all five isochrones to the clusters gives
us reasonable confidence to interpolate our corrections. Smoothly
changing corrections can be applied to the entire set of isochrones,
then fit to an unknown population, although caution must be taken when
extending beyond our calibrating clusters, i.e. M92 at \textrm{[Fe/H]} $=
-2.3$, and NGC 6791 at \textrm{[Fe/H]} $= +0.40$. For the majority of cluster
CMDs the empirical corrections improve the isochrone-ridgeline
alignment such that the average difference between the two is
$\lesssim$ 0.01 mag along the giant branch. In Figures \ref{CMDs} -
\ref{medCMD} the dash-dot lines represent the uncorrected isochrones,
while the solid lines are the empirically corrected isochrones.

All the color corrections listed in Table 8 are shown to \textrm{[Fe/H]} $=-2.5
$.  Below this metallicity we replace all of the metallicity dependent
terms with the value of the term when \textrm{[Fe/H]} $= -2.5$ keeping the
corrections constant where the color change is small and we are unable
to verify the corrections.

\subsection{Metallicity Determinations}
We tested the empirical isochrone corrections by re-deriving the
cluster metallicities. We assign photometric metallicities to stars in
the CMDs by searching isochrone grids with spacing in \textrm{[Fe/H]}
of 0.05 dex; each star was assigned the metallicity of the closest
isochrone.  We adopted [$\alpha$/Fe] $ = 0.0$ since we cannot separate
the  color effects due to $\alpha$ and \textrm{[Fe/H]} with only three 
filters.  We selected giant branch stars with errors $<$0.03 mag in
all filters.

\subsubsection{Metallicities From CMDs}

The photometric metallicities adopted from (F336W--F555W, F814W),
(F390M--F555W, F814W) and (F390W--F555W, F814W) were used to create
MDFs for every cluster, which are shown on the left side of Figure
\ref{plotmdf}.  For all the clusters the MDF peaks are within $\pm$
0.06 dex of the spectroscopically derived values. The
recovered peaks, widths and the number of stars measured are listed in
Table 9.

\begin{figure}[ht]
  \epsfig{file=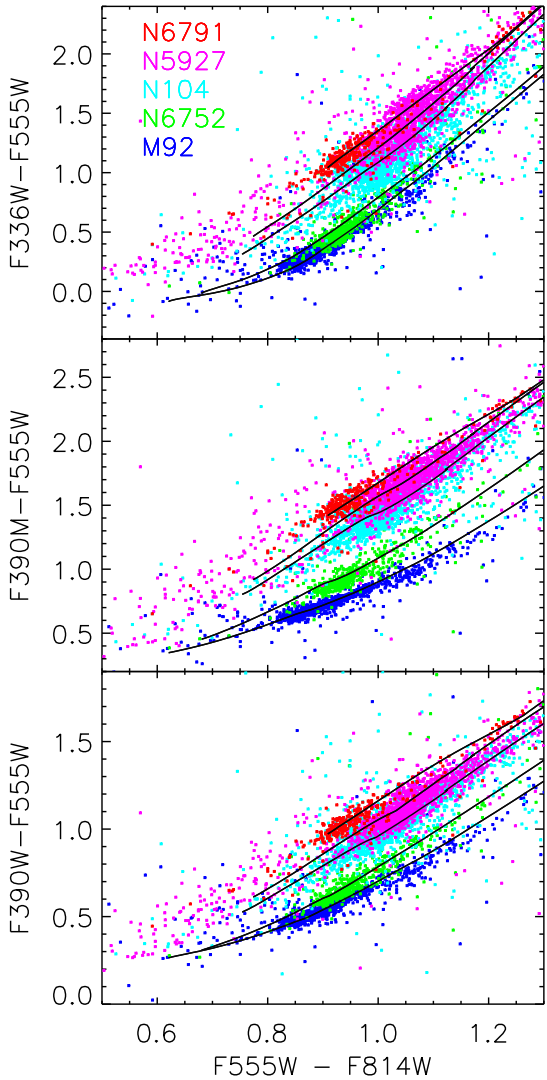,height=6in }
  \caption{Shown above are the (F336W--F555W, F555W--F814W),
    (F390M--F555W, F555W--F814W),  and (F390W--F555W, F555W--F814W)
    color-color diagrams of the five  clusters with corrected
    literature valued isochrones overlaid in  black.  The corrected
    isochrones are available in the online version of Table 10.  (A
    color version of this figure is  available in the  online journal.) 	\label{cc}}  
\end{figure}

The MDF dispersion within each cluster does not vary much across the
metallicity range, although we do see slightly larger numbers for the 
differentially reddened NGC 5927 and for the lowest metallicity
cluster, M92, reflecting the diminished accuracy expected at low
metallicity, as discussed in Section 2.2. Additionally for NGC 104 and
NGC 5927 the horizontal branch stars are clearly seen as small
shoulders left of the MDF peak.  The horizontal branch metallicities
are systematically lower than the true metallicities because we only
use the isochrone colors along the giant branch to assign metallicity,
so horizontal branch stars were incorrectly matched to GB stars.
Overall, the average MDF dispersion for all clusters is $\sim$0.10 dex
and is consistent with what is expected from photometric errors.  We
simulated this by creating synthetic CMDs based on the IMF and 
photometric errors within our data set. At each magnitude along the GB
ridgeline we used the measured photometric errors to randomly
distribute the relative number of stars found at each magnitude.  The
synthetic CMD was then put through the same isochrone matching
processes to get photometric metallicities and a MDF for each
cluster.

In order to determine the systematic effect of using uncorrected
isochrones to assign photometric metallicities we compared the
spectroscopically determined metallicities to those adopted using the
uncorrected isochrone grid to assign metallicities.  Consistent
cluster metallicities could not be found across the various CMDs. For
a given CMD the peak of the metallicity distribution varied by $\pm$
0.4 dex from the canonical values, with the larger discrepancies
occurring at low metallicity.

\begin{deluxetable}{cllll}[H]
\tabletypesize{\tiny}
\tablenum{10}
\tablecaption{Corrected isochrone sequence in Color-Color Diagrams}
\tablehead{
	\colhead{[Fe/H]}  &
	\colhead{(F555W--} &
   	\colhead{(F336W--} &
   	\colhead{(F390M--} &
 	\colhead{(F390W--} \\   
		\colhead{} &
  		\colhead{F814W)} &   
		\colhead{F555W)} &
   		\colhead{F555W)} &
   		\colhead{F555W)} }
\startdata
-2.30 &  0.6208& -0.0853&  0.3473&  0.2647	\\
-2.30 &  0.6212& -0.0839&  0.3476&  0.2651	\\
-2.30 &  0.6217& -0.0825&  0.3480&  0.2657	\\
. . . .&. . . .&. . . .&. . . .&. . . .\\
-1.45 &  0.6804& -0.0160&  0.4533&  0.3075   \\
-1.45 &  0.6807& -0.0149&  0.4534&  0.3077   \\
-1.45 &  0.6809& -0.0137&  0.4536&  0.3080   \\
. . . .&. . . .&. . . .&. . . . &. . . .\\
-0.70 &  0.7542&  0.3142&  0.8049&  0.5268   \\ 
-0.70 &  0.7545&  0.3152&  0.8052&  0.5270   \\ 
-0.70 &  0.7550&  0.3165&  0.8058&  0.5276   \\ 
. . . .&. . . .&. . . .&. . . .&. . . .\\
-0.40 &  0.7740&  0.4647&  0.9201&  0.6125   \\  
-0.40 &  0.7743&  0.4656&  0.9203&  0.6129   \\  
-0.40 &  0.7746&  0.4668&  0.9209&  0.6133   \\  
. . . .&. . . .&. . . .&. . . . &. . . .\\
+0.40 &  0.9087&  1.0424&  1.4213&  0.9728\\
+0.40 &  0.9089&  1.0435&  1.4217&  0.9731\\
+0.40 &  0.9094&  1.0451&  1.4223&  0.9737
\enddata
\tablecomments{Table 10 is published in its entirety in the electronic
  edition of AJ. A portion is shown here for guidance regarding its
  form and content.\label{ccdat}}
\end{deluxetable}

\subsubsection{Metallicities From Color-Color Diagrams}

To test our metallicity recovery without using age as a known
parameter, we apply the empirical corrections to isochrones in
color-color diagrams (Figure \ref{cc}) then adopt metallicities. For
each star we found the closest isochrone in the grid (spaced by 0.05
dex in \textrm{[Fe/H]} with the [$\alpha$/Fe] held constant at solar) and
assigned the corresponding metallicity. The corrected fiducial
isochrone colors shown as black lines in the color-color diagram of
Figure \ref{cc} are available in the online version in Table 10.  
The MDFs derived from color-color diagrams are shown on the right side
of Figure \ref{plotmdf}.  The recovered peaks, widths and the number
of stars measured are listed in Table 11. We selected the
same giant branch stars as before, i.e. stars with photometry better
than 0.03 mag in all filters. We exclude poorly fit stars, e.g. stars
more than 0.25 mag outside of the color range covered by the
isochrone grid.  This additional constraint accounts for the
varying number of stars measured for different color-color diagrams in
the last column of Table 11, though the variation is $\sim 1\%$.

For four of the five clusters, only the giant branch stars are used to
adopt metallicities in the color-color plots. For the fifth cluster,
NGC 6791, there are not enough prominent giant branch stars,
necessitating that we additionally use the sub-giant branch and main
sequence stars. Theoretically the different evolutionary states should 
not affect the metallicity determination because we selected the
corresponding portion of the isochrone to match the less evolved
stars.

The MDFs from color-color diagrams, as compared to those from CMDs,
tend to have slightly larger offsets from the spectroscopically
derived values, and systematically larger MDF dispersions. However,
these metallicities were adopted without any age information, which is
necessary for measuring metallicities for a population of mixed ages.
The MDFs from the (F390M--F555W) color-color diagrams have mean
metallicities within $\pm$ 0.07 dex for all clusters.  The
MDFs from the (F390W--F555W) color-color diagrams produces mean metallicities
within $\pm$ 0.05 dex for M92, NGC 104 (47Tuc) and NGC 6791, and peaks within
$\sim$ 0.2 dex for NGC 6752 and the differentially reddened NGC 5927.
The MDFs from the (F336W--F555W) color-color diagram finds mean
metallicities of $\sim$0.05 dex for most clusters, with the exception of
NGC 104 which is discrepant by 0.1 dex.

The larger MDF widths from the color-color diagrams as opposed to the
CMD are a natural consequence of using color on the y-axis as opposed
to magnitude. The change in color for the total range in metallicity
is generally smaller than that for magnitude (see Section 2), thus
there is decreased resolution when using color-color diagrams;
additionally, the errors for colors are larger than for magnitudes.
Together, the decreased color span and the increased errors both add
to the uncertainty in the photometric metallicities adopted from
color-color diagrams widening the MDFs. The derived MDF widths are
consistent with what is expected based upon the photometric errors.

\begin{deluxetable}{lllll}[Hb]
\tabletypesize{\scriptsize}
\tablenum{11}
\tablecaption{Cluster MDF Parameters from CC diagrams     }
\tablehead{
  \colhead{Color Color}     &
  \colhead{Cluster} &
  \colhead{Peak}    &
  \colhead{$\sigma$}    &
  \colhead{$\#$ of } \\ 
     \colhead{Diagram} &
     \colhead{} &
     \colhead{\textrm{[Fe/H]}} &
     \colhead{} &
     \colhead{stars}  }

\startdata
(F336W--F555W,       &       M92 &  -2.22  & 0.471  &     879  \\   
 F555W--F814W)       &     N6752 &  -1.73  & 0.484  &     454  \\   
 . . . . . . . . . . &      N104 &  -0.79  & 0.285  &    2067  \\   
 . . . . . . . . . . &     N5927 &  -0.38  & 0.365  &    1147  \\   
 . . . . . . . . . . &     N6791 &   0.37  & 0.075  &     320  \\   
\hline  			                                                 
(F390M--F555W,       &       M92 &  -2.25  & 0.244  &     906  \\   
 F555W--F814W)       &     N6752 &  -1.46  & 0.172  &     448  \\   
 . . . . . . . . . . &      N104 &  -0.68  & 0.201  &    2110  \\  
 . . . . . . . . . . &     N5927 &  -0.47  & 0.264  &    1322  \\   
 . . . . . . . . . . &     N6791 &   0.37  & 0.061  &     314  \\   
\hline                                                        
(F390W--F555W,       &       M92 &  -2.27  & 0.396  &     891  \\   
 F555W--F814W)       &     N6752 &  -1.57  & 0.250  &     447  \\   
 . . . . . . . . . . &      N104 &  -0.70  & 0.232  &    2107  \\   
 . . . . . . . . . . &     N5927 &  -0.58  & 0.296  &    1298  \\   
 . . . . . . . . . . &     N6791 &   0.36  & 0.073  &     303     
\enddata
\label{mdfcc}
\end{deluxetable}

Of the three color-color diagrams tested to assign metallicity the
(F336W--F555W, F555W--F814W) diagram does the worst. For NGC 104
the peak metallicity is off by 0.09 dex and the MDF dispersions are
twice to three times the dispersions from (F390M--F555W, F555W--F814W)
depending on the cluster. F336W, while predicted to remain sensitive
at low metallicity, does not show this in practice, as seen from the
wide dispersions for the clusters. The overestimate of the sensitivity
could be due to the fact that the color separation, as seen from the
clusters, is smaller than predicted from the isochrones, both of which
are shown in Figure \ref{uCMD}.

The (F390M--F555W, F555W--F814W) color-color diagram generally does
the best job deriving metallicities of all the colors.  The
MDFs have peaks within $\pm$ 0.07 dex of the spectroscopically
determined metallicities for all clusters.  The MDF dispersions are
$\lesssim$ 0.2 dex, except for the very-low metallicity M92 at 0.24 dex
and the differentially reddened NGC 5927 at 0.26 dex. The MDFs derived
using the (F390M--F555W) color are better than those from
(F390W--F555W) and (F336W--F555W).  In particular the MDF dispersion in the
very-metal poor regime is the narrowest of all the MDFs from
color-color diagrams by almost a factor of two. The greater 
accuracy in photometric metallicities below \textrm{[Fe/H]} $<-1.0$ when using
the F390M makes this filter the most useful when working in the
very-low metallicity regime.

The (F390W--F555W, F555W--F814W) color-color diagram recovers peaks
within $\pm$ 0.2 dex of the spectroscopically determined
metallicities.  The MDF dispersions are $\sim$ 0.25 dex, except for 
M92, the very-low metallicity cluster, where the dispersion is 0.4
dex.  However, F390W is very economical in terms of integration time,
and accurate to $\sim$0.25 dex above \textrm{[Fe/H]} $> -2.0$,
therefore this filter is most useful when working in the more metal
rich regimes.

In order to determine the systematic effect of using uncorrected
isochrones to assign photometric metallicities we compared the
spectroscopically determined metallicities to those adopted using the 
uncorrected isochrone grid in color-color diagrams to assign
metallicities. In the MDFs from the color-color diagrams of
(F336W--F555W, F555W--F814W) and (F390W--F555W, F555W--F814W) the
adopted cluster metallicities were within $\sim$0.2 dex of the
spectroscopically derived values for the clusters with metallicities
between  $-1.5 <$ \textrm{[Fe/H]} $< 0$, while the very metal poor and
metal rich cluster are 3 to 4 times more discrepant. In the
color-color diagram of (F390M--F555W, F555W--F814W) the metallicities
are all systematically higher by 0.4 to 0.8 dex, again, with more
disparate values for very metal poor and metal rich clusters.

\section{Conclusion}

We explored the metallicity and temperature sensitivities of colors
created from nine WFC3/UVIS filters aboard the HST using Dartmouth
isochrones and Kurucz atmospheres models. The theoretical isochrone
colors were tested and calibrated against observations of five well
studied clusters. 

We found that (F390W--F555W), and  (F390M--F555W), are the most
promising colors in terms of metallicity sensitivity.  For almost all
of the clusters F390M has slightly better metallicity sensitivity, and
narrower MDF dispersions, although the F390W filter requires much less
integration time. Additionally, at low metallicity the photometric
metallicities from (F390M--F555W) are nearly twice as accurate as
those from (F390W--F555W).

Using photometry of M92, NGC 6752, NGC 104, NGC 5927 and NGC 6791, all
of which have spectroscopically determined metallicities spanning
$-2.30 <$ \textrm{[Fe/H]} $< +0.4$, we found empirical corrections to the
Dartmouth isochrone grid for each of the following CMDs
(F555W--F814W, F814W), (F336W--F555W, F814W), (F390M--F555W, F814W) 
and (F390W--F555W, F814W).

Using the empirical corrections we tested the accuracy and spread of
the photometric metallicities adopted from CMDs and color-color
diagrams.  From the color-color diagrams we were able to recover the
spectroscopic metallicities independent from any assumptions about
cluster ages, which allows us to apply the color-color diagram method
of determining metallicities to complex stellar populations with
confidence that the method breaks the age-metallicity degeneracy.

When using color-color diagrams to assign metallicity, we
found the (F390M--F555W) color to have the greatest accuracy and
consistency across the entire metallicity range, with the main
advantage being the increased sensitivity at low metallicity, while
(F336W--F555W) and (F390W--F555W) both lose accuracy in this range. We
showed that by using the calibrated isochrones we could assign the
overall cluster metallicity to within $\sim$ 0.1 dex in \textrm{[Fe/H]} when
using CMDs (i.e when the distance, reddening and ages are
approximately known).  The measured MDFs from color-color diagrams
show this method measures metallicities of stellar clusters of unknown
age and metallicity with an accuracy of $\sim$ 0.2 - 0.58 dex using
F336W, $\sim$0.15 - 0.3 dex using F390M, and $\sim$0.2 - 0.52 dex with
F390W, with the larger uncertainty pertaining to lowest metallicity
range. 

\section{Acknowledgements}
Support for program 11729 was provided
by NASA through a grant from the Space Telescope Science Institute,
which is operated by the Association of Universities for Research in
Astronomy, Inc., under NASA contract NAS 5-26555.

\begin{turnpage}

\begin{deluxetable*}{ll}[H]
\tablenum{8}
\tabletypesize{\scriptsize}
\tablecaption{Empirical corrections for isochrone colors}
\tablehead{
  \colhead{Correction}&
  \colhead{Conditions}
}

\startdata
$\rm{(F555W-F814W)}_{n}=(0.89+0.028\rm{\textrm{[Fe/H]}})
\rm{(F555W-F814W)}_{o} + 0.136  $ 
& log g $>$ 3.4, $\&$ -2.5 $\leq$ \textrm{[Fe/H]}$<$ -1.4\\ 

$\rm{(F555W-F814W)}_{n}=(0.89+0.028\rm{\textrm{[Fe/H]}})
\rm{(F555W-F814W)}_{o} + 0.136
- (0.015\rm{\textrm{[Fe/H]}})(3.4-\log{g})$
&log g $<$ 3.4, $\&$ -2.5 $\leq$ \textrm{[Fe/H]}$<$ -1.4\\

$\rm{(F555W-F814W)}_{n}=(0.89+0.028\rm{\textrm{[Fe/H]}})
\rm{(F555W-F814W)}_{o} +
0.05(\rm{\textrm{[Fe/H]}}+0.55)^2+0.10 $
&log g $>$ 3.4, $\&$ -1.4 $\leq$ \textrm{[Fe/H]}$<$ 0.5 \\  

$\rm{(F555W-F814W)}_{n}=(0.89+0.028\rm{\textrm{[Fe/H]}})
\rm{(F555W-F814W)}_{o}+0.05(\rm{\textrm{[Fe/H]}}+0.55)^2
+0.10-(0.015\rm{\textrm{[Fe/H]}})(3.4-\log{g})$
&log g $<$ 3.4, $\&$ -1.4 $\leq$ \textrm{[Fe/H]}$<$ 0.5 \\  
\\
\hline

$\rm{(F336W-F555W)}_{n}=0.88\rm{(F336W-F555W)}_{o} - 0.009 $
& log g $>$ 3.4, $\&$ -2.5 $<$ \textrm{[Fe/H]}$\leq$ -1.65 \\

$\rm{(F336W-F555W)}_{n}=0.88\rm{(F336W-F555W)}_{o} - 0.009 
+ (0.015-0.036\rm{\textrm{[Fe/H]}})(3.4-\log{g})$
&log g $<$ 3.4, $\&$ -2.5 $<$ \textrm{[Fe/H]}$\leq$ -1.65 \\

$\rm{(F336W-F555W)}_{n}=0.88\rm{(F336W-F555W)}_{o} + 
0.04(\rm{\textrm{[Fe/H]}}+0.80)^2-0.02 $
& log g $>$ 3.4, $\&$ -1.65$<$\textrm{[Fe/H]} $\leq$ 0.05\\ 

$\rm{(F336W-F555W)}_{n}=0.88\rm{(F336W-F555W)}_{o} +
0.04(\rm{\textrm{[Fe/H]}}+0.80)^2 -0.02 +
(0.015-0.036\rm{\textrm{[Fe/H]}}(3.4-\log{g}) $
&log g $<$ 3.4, $\&$ -1.65$<$\textrm{[Fe/H]} $\leq$ 0.05\\

$\rm{(F336W-F555W)}_{n}=0.88\rm{(F336W-F555W)}_{o}- 0.009 $
& log g $>$ 3.4, $\&$ 0.05$<$\textrm{[Fe/H]} $\leq$ 0.5\\

$\rm{(F336W-F555W)}_{n}=0.88\rm{(F336W-F555W)}_{o} - 0.009
 +(0.015-0.036\rm{\textrm{[Fe/H]}})(3.4-\log{g}) $
&log g $<$ 3.4, $\&$ 0.05$<$\textrm{[Fe/H]} $\leq$ 0.5\\ 
\hline
\\

$\rm{(F390W-F555W)}_{n}=(0.89+0.012\rm{\textrm{[Fe/H]}})
\rm{(F390W-F555W)}_{o}+0.01(\rm{\textrm{[Fe/H]}}+2.5)^2 + 0.05 $
&log g $>$ 3.3, $\&$ -2.5 $<$ \textrm{[Fe/H]} $<$ 1.65 \\  

$\rm{(F390W-F555W)}_{n}=(0.89+0.012\rm{\textrm{[Fe/H]}})
\rm{(F390W-F555W)}_{o}+0.01(\rm{\textrm{[Fe/H]}}+2.5)^2 + 0.05 +
(0.07 +0.006\rm{\textrm{[Fe/H]}})(3.3-\log{g}) $
&log g $<$ 3.3, $\&$ -2.5 $<$ \textrm{[Fe/H]} $<$ 1.65 \\

$\rm{(F390W-F555W)}_{n}=(0.89+0.012\rm{\textrm{[Fe/H]}})
\rm{(F390W-F555W)}_{o}+0.057 $
&log g $>$ 3.3, $\&$ -1.65 $<$ \textrm{[Fe/H]} $<$ +0.5 \\  

$\rm{(F390W-F555W)}_{n}=(0.89+0.012\rm{\textrm{[Fe/H]}})
\rm{(F390W-F555W)}_{o}+0.057 +
(0.07 +0.006\rm{\textrm{[Fe/H]}})(3.3-\log{g}) $
&log g $<$ 3.3, $\&$ -1.65 $<$ \textrm{[Fe/H]} $<$ 0.5 \\  

\hline
\\

$\rm{(F390M-F555W)}_{n}=(1.01-0.005\rm{\textrm{[Fe/H]}})
\rm{(F390M-F555W)}_{o} +0.07(\rm{\textrm{[Fe/H]}}+2.50)^2+0.005 $
&log g $>$ 3.2, $\&$ -2.5 $<$ \textrm{[Fe/H]}$<$  -1.65 \\

$\rm{(F390M-F555W)}_{n}=(1.01-0.005\rm{\textrm{[Fe/H]}})
\rm{(F390M-F555W)}_{o} +0.07(\rm{\textrm{[Fe/H]}}+2.50)^2+0.005
+(0.018-0.011\rm{\textrm{[Fe/H]}})(3.2-\log{g})$ 
& log g $<$ 3.2, $\&$ -2.5 $<$  \textrm{[Fe/H]}$<$ -1.65 \\

$\rm{(F390M-F555W)}_{n}=(1.01-0.005\rm{\textrm{[Fe/H]}})
\rm{(F390M-F555W)}_{o}+0.056$
&log g $>$ 3.2, $\&$ -1.65 $<$ \textrm{[Fe/H]}$<$  +0.5 \\

$\rm{(F390M-F555W)}_{n}=(1.01-0.005\rm{\textrm{[Fe/H]}})
\rm{(F390M-F555W)}_{o}+0.056 +(0.018-0.011
\rm{\textrm{[Fe/H]}})(3.2-\log{g})$ 
& log g $<$ 3.2, $\&$ -1.65 $<$ \textrm{[Fe/H]}$<$+0.5
\enddata
\tablecomments{Derived isochrone corrections. \label{correction}}
\end{deluxetable*}

\end{turnpage}


\begin{thebibliography}{49}
\expandafter\ifx\csname natexlab\endcsname\relax\def\natexlab#1{#1}\fi

\bibitem[{{Anthony-Twarog} {et~al.}(1991){Anthony-Twarog}, {Twarog}, {Laird},
  \& {Payne}}]{1991AJ....101.1902A}
{Anthony-Twarog}, B.~J., {Twarog}, B.~A., {Laird}, J.~B., \& {Payne}, D. 1991,
  \aj, 101, 1902

\bibitem[{{Armandroff} \& {Zinn}(1988)}]{1988AJ.....96...92A}
{Armandroff}, T.~E., \& {Zinn}, R. 1988, \aj, 96, 92

\bibitem[{{Bonatto} {et~al.}(2013){Bonatto}, {Campos}, \&
  {Kepler}}]{2013MmSAI..84..187B}
{Bonatto}, C., {Campos}, F., \& {Kepler}, S.~O. 2013, \memsai, 84, 187

\bibitem[{{Brogaard} {et~al.}(2012){Brogaard}, {VandenBerg}, {Bruntt},
  {Grundahl}, {Frandsen}, {Bedin}, {Milone}, {Dotter}, {Feiden}, {Stetson},
  {Sandquist}, {Miglio}, {Stello}, \& {Jessen-Hansen}}]{2012A&A...543A.106B}
{Brogaard}, K., {VandenBerg}, D.~A., {Bruntt}, H., {et~al.} 2012, \aap, 543,
  A106

\bibitem[{{Brown}(2008)}]{2008hst..prop11664B}
{Brown}, T. 2008, in HST Proposal, 11664

\bibitem[{{Brown} {et~al.}(2005){Brown}, {Ferguson}, {Smith}, {Guhathakurta},
  {Kimble}, {Sweigart}, {Renzini}, {Rich}, \&
  {VandenBerg}}]{2005AJ....130.1693B}
{Brown}, T.~M., {Ferguson}, H.~C., {Smith}, E., {et~al.} 2005, \aj, 130, 1693

\bibitem[{{Brown} {et~al.}(2009){Brown}, {Sahu}, {Zoccali}, {Renzini},
  {Ferguson}, {Anderson}, {Smith}, {Bond}, {Minniti}, {Valenti}, {Casertano},
  {Livio}, {Panagia}, {Vanden Berg}, \& {Valenti}}]{2009AJ....137.3172B}
{Brown}, T.~M., {Sahu}, K., {Zoccali}, M., {et~al.} 2009, \aj, 137, 3172

\bibitem[{{Calamida} {et~al.}(2007){Calamida}, {Bono}, {Stetson}, {Freyhammer},
  {Cassisi}, {Grundahl}, {Pietrinferni}, {Hilker}, {Primas}, {Richtler},
  {Romaniello}, {Buonanno}, {Caputo}, {Castellani}, {Corsi}, {Ferraro},
  {Iannicola}, \& {Pulone}}]{2007ApJ...670..400C}
{Calamida}, A., {Bono}, G., {Stetson}, P.~B., {et~al.} 2007, \apj, 670, 400

\bibitem[{{Canterna}(1976)}]{1976AJ.....81..228C}
{Canterna}, R. 1976, \aj, 81, 228

\bibitem[{{Carraro} {et~al.}(2006){Carraro}, {Villanova}, {Demarque},
  {McSwain}, {Piotto}, \& {Bedin}}]{2006ApJ...643.1151C}
{Carraro}, G., {Villanova}, S., {Demarque}, P., {et~al.} 2006, \apj, 643, 1151

\bibitem[{{Carretta} {et~al.}(2009){Carretta}, {Bragaglia}, {Gratton},
  {D'Orazi}, \& {Lucatello}}]{2009A&A...508..695C}
{Carretta}, E., {Bragaglia}, A., {Gratton}, R., {D'Orazi}, V., \& {Lucatello},
  S. 2009, \aap, 508, 695

\bibitem[{{Carretta} \& {Gratton}(1997)}]{1997A&AS..121...95C}
{Carretta}, E., \& {Gratton}, R.~G. 1997, \aaps, 121, 95

\bibitem[{{Castelli} \& {Kurucz}(2004)}]{2004astro.ph..5087C}
{Castelli}, F., \& {Kurucz}, R.~L. 2004, ArXiv Astrophysics e-prints

\bibitem[{{Chaboyer} {et~al.}(1999){Chaboyer}, {Green}, \&
  {Liebert}}]{1999AJ....117.1360C}
{Chaboyer}, B., {Green}, E.~M., \& {Liebert}, J. 1999, \aj, 117, 1360

\bibitem[{{Da Costa} \& {Armandroff}(1990)}]{1990AJ....100..162D}
{Da Costa}, G.~S., \& {Armandroff}, T.~E. 1990, \aj, 100, 162

\bibitem[{{Da Costa} {et~al.}(2002){Da Costa}, {Armandroff}, \&
  {Caldwell}}]{2002AJ....124..332D}
{Da Costa}, G.~S., {Armandroff}, T.~E., \& {Caldwell}, N. 2002, \aj, 124, 332

\bibitem[{{Da Costa} {et~al.}(2000){Da Costa}, {Armandroff}, {Caldwell}, \&
  {Seitzer}}]{2000AJ....119..705D}
{Da Costa}, G.~S., {Armandroff}, T.~E., {Caldwell}, N., \& {Seitzer}, P. 2000,
  \aj, 119, 705

\bibitem[{{De Angeli} {et~al.}(2005){De Angeli}, {Piotto}, {Cassisi}, {Busso},
  {Recio-Blanco}, {Salaris}, {Aparicio}, \& {Rosenberg}}]{2005AJ....130..116D}
{De Angeli}, F., {Piotto}, G., {Cassisi}, S., {et~al.} 2005, \aj, 130, 116

\bibitem[{{di Cecco} {et~al.}(2010){di Cecco}, {Becucci}, {Bono}, {Monelli},
  {Stetson}, {Degl'Innocenti}, {Prada Moroni}, {Nonino}, {Weiss}, {Buonanno},
  {Calamida}, {Caputo}, {Corsi}, {Ferraro}, {Iannicola}, {Pulone},
  {Romaniello}, \& {Walker}}]{2010PASP..122..991D}
{di Cecco}, A., {Becucci}, R., {Bono}, G., {et~al.} 2010, \pasp, 122, 991

\bibitem[{{Dotter} {et~al.}(2008){Dotter}, {Chaboyer}, {Jevremovi{\'c}},
  {Kostov}, {Baron}, \& {Ferguson}}]{2008ApJS..178...89D}
{Dotter}, A., {Chaboyer}, B., {Jevremovi{\'c}}, D., {et~al.} 2008, \apjs, 178,
  89

\bibitem[{{Gallart}(2008)}]{2008ASPC..390..278G}
{Gallart}, C. 2008, in Astronomical Society of the Pacific Conference Series,
  Vol. 390, Pathways Through an Eclectic Universe, ed. J.~H. {Knapen}, T.~J.
  {Mahoney}, \& A.~{Vazdekis}, 278

\bibitem[{{Garc{\'{\i}}a-Berro} {et~al.}(2013){Garc{\'{\i}}a-Berro}, {Torres},
  {Isern}, {Salaris}, {C{\'o}rsico}, \& {Althaus}}]{2013EPJWC..4305003G}
{Garc{\'{\i}}a-Berro}, E., {Torres}, S., {Isern}, J., {et~al.} 2013, in
  European Physical Journal Web of Conferences, Vol.~43, European Physical
  Journal Web of Conferences, 5003

\bibitem[{{Geisler} {et~al.}(2012){Geisler}, {Villanova}, {Carraro},
  {Pilachowski}, {Cummings}, {Johnson}, \& {Bresolin}}]{2012arXiv1207.3328G}
{Geisler}, D., {Villanova}, S., {Carraro}, G., {et~al.} 2012, ArXiv e-prints

\bibitem[{{Gratton} {et~al.}(2003){Gratton}, {Bragaglia}, {Carretta},
  {Clementini}, {Desidera}, {Grundahl}, \& {Lucatello}}]{2003A&A...408..529G}
{Gratton}, R.~G., {Bragaglia}, A., {Carretta}, E., {et~al.} 2003, \aap, 408,
  529

\bibitem[{{Gratton} {et~al.}(2005){Gratton}, {Bragaglia}, {Carretta}, {de
  Angeli}, {Lucatello}, {Piotto}, \& {Recio Blanco}}]{2005A&A...440..901G}
---. 2005, \aap, 440, 901

\bibitem[{{Gratton} {et~al.}(2012){Gratton}, {Carretta}, \&
  {Bragaglia}}]{2012A&ARv..20...50G}
{Gratton}, R.~G., {Carretta}, E., \& {Bragaglia}, A. 2012, \aapr, 20, 50

\bibitem[{{Grundahl} {et~al.}(2008){Grundahl}, {Clausen}, {Hardis}, \&
  {Frandsen}}]{2008A&A...492..171G}
{Grundahl}, F., {Clausen}, J.~V., {Hardis}, S., \& {Frandsen}, S. 2008, \aap,
  492, 171

\bibitem[{{Harris}(1996)}]{1996AJ....112.1487H}
{Harris}, W.~E. 1996, \aj, 112, 1487

\bibitem[{{Heitsch} \& {Richtler}(1999)}]{1999A&A...347..455H}
{Heitsch}, F., \& {Richtler}, T. 1999, \aap, 347, 455

\bibitem[{{Holtzman}(2008)}]{2008hst..prop11729H}
{Holtzman}, J. 2008, in HST Proposal, 11729

\bibitem[{{Holtzman}(2009)}]{2009hst..prop12304H}
{Holtzman}, J. 2009, in HST Proposal, 12304

\bibitem[{{Johnson} \& {Morgan}(1953)}]{1953ApJ...117..313J}
{Johnson}, H.~L., \& {Morgan}, W.~W. 1953, \apj, 117, 313

\bibitem[{{Kirby} {et~al.}(2011){Kirby}, {Lanfranchi}, {Simon}, {Cohen}, \&
  {Guhathakurta}}]{2011ApJ...727...78K}
{Kirby}, E.~N., {Lanfranchi}, G.~A., {Simon}, J.~D., {Cohen}, J.~G., \&
  {Guhathakurta}, P. 2011, \apj, 727, 78

\bibitem[{{Kraft} \& {Ivans}(2003)}]{2003PASP..115..143K}
{Kraft}, R.~P., \& {Ivans}, I.~I. 2003, \pasp, 115, 143

\bibitem[{{Lianou} {et~al.}(2011){Lianou}, {Grebel}, \&
  {Koch}}]{2011A&A...531A.152L}
{Lianou}, S., {Grebel}, E.~K., \& {Koch}, A. 2011, \aap, 531, A152

\bibitem[{{Mar{\'{\i}}n-Franch} {et~al.}(2009){Mar{\'{\i}}n-Franch},
  {Aparicio}, {Piotto}, {Rosenberg}, {Chaboyer}, {Sarajedini}, {Siegel},
  {Anderson}, {Bedin}, {Dotter}, {Hempel}, {King}, {Majewski}, {Milone},
  {Paust}, \& {Reid}}]{2009ApJ...694.1498M}
{Mar{\'{\i}}n-Franch}, A., {Aparicio}, A., {Piotto}, G., {et~al.} 2009, \apj,
  694, 1498

\bibitem[{{McWilliam}(2010)}]{2010nuco.confE...8M}
{McWilliam}, A. 2010, in Nuclei in the Cosmos

\bibitem[{{Origlia} {et~al.}(2006){Origlia}, {Valenti}, {Rich}, \&
  {Ferraro}}]{2006ApJ...646..499O}
{Origlia}, L., {Valenti}, E., {Rich}, R.~M., \& {Ferraro}, F.~R. 2006, \apj,
  646, 499

\bibitem[{{Renzini} {et~al.}(1996){Renzini}, {Bragaglia}, {Ferraro},
  {Gilmozzi}, {Ortolani}, {Holberg}, {Liebert}, {Wesemael}, \&
  {Bohlin}}]{1996ApJ...465L..23R}
{Renzini}, A., {Bragaglia}, A., {Ferraro}, F.~R., {et~al.} 1996, \apjl, 465,
  L23

\bibitem[{{Saviane} {et~al.}(2000){Saviane}, {Rosenberg}, {Piotto}, \&
  {Aparicio}}]{2000A&A...355..966S}
{Saviane}, I., {Rosenberg}, A., {Piotto}, G., \& {Aparicio}, A. 2000, \aap,
  355, 966

\bibitem[{{Str{\"o}mgren}(1966)}]{1966ARA&A...4..433S}
{Str{\"o}mgren}, B. 1966, \araa, 4, 433

\bibitem[{{Tolstoy} {et~al.}(2009){Tolstoy}, {Hill}, \&
  {Tosi}}]{2009ARA&A..47..371T}
{Tolstoy}, E., {Hill}, V., \& {Tosi}, M. 2009, \araa, 47, 371

\bibitem[{{VandenBerg}(2000)}]{2000ApJS..129..315V}
{VandenBerg}, D.~A. 2000, \apjs, 129, 315

\bibitem[{{VandenBerg} {et~al.}(2012){VandenBerg}, {Bergbusch}, {Dotter},
  {Ferguson}, {Michaud}, {Richer}, \& {Proffitt}}]{2012ApJ...755...15V}
{VandenBerg}, D.~A., {Bergbusch}, P.~A., {Dotter}, A., {et~al.} 2012, \apj,
  755, 15

\bibitem[{{VandenBerg} \& {Clem}(2003)}]{2003AJ....126..778V}
{VandenBerg}, D.~A., \& {Clem}, J.~L. 2003, \aj, 126, 778

\bibitem[{{Woodley} {et~al.}(2012){Woodley}, {Goldsbury}, {Kalirai}, {Richer},
  {Tremblay}, {Anderson}, {Bergeron}, {Dotter}, {Esteves}, {Fahlman}, {Hansen},
  {Heyl}, {Hurley}, {Rich}, {Shara}, \& {Stetson}}]{2012AJ....143...50W}
{Woodley}, K.~A., {Goldsbury}, R., {Kalirai}, J.~S., {et~al.} 2012, \aj, 143,
  50

\bibitem[{{Wylie} {et~al.}(2006){Wylie}, {Cottrell}, {Sneden}, \&
  {Lattanzio}}]{2006ApJ...649..248W}
{Wylie}, E.~C., {Cottrell}, P.~L., {Sneden}, C.~A., \& {Lattanzio}, J.~C. 2006,
  \apj, 649, 248

\bibitem[{{Zinn} \& {West}(1984)}]{1984ApJS...55...45Z}
{Zinn}, R., \& {West}, M.~J. 1984, \apjs, 55, 45

\bibitem[{{Zoccali} {et~al.}(2001){Zoccali}, {Renzini}, {Ortolani},
  {Bragaglia}, {Bohlin}, {Carretta}, {Ferraro}, {Gilmozzi}, {Holberg},
  {Marconi}, {Rich}, \& {Wesemael}}]{2001ApJ...553..733Z}
{Zoccali}, M., {Renzini}, A., {Ortolani}, S., {et~al.} 2001, \apj, 553, 733

\end{thebibliography}
\end{document}